\documentclass[a4paper,UKenglish,cleveref, autoref, thm-restate]{lipics-v2021}

\usepackage{amsthm}
\usepackage{dsfont}

\usepackage{todonotes}
\usepackage{varwidth}
\usepackage{bbold}
\usepackage{xcolor}

\usepackage[normalem]{ulem}
\usepackage{xspace}
\usepackage{mdframed}
\usepackage{wasysym}
\usepackage{paralist}

\usepackage{graphicx} 
\usepackage{caption}
\usepackage{epstopdf}
\usepackage{float}

\usepackage[ruled,vlined,linesnumbered]{algorithm2e}
\usepackage{setspace}

\usepackage{tikz}
\usetikzlibrary{arrows,decorations.pathmorphing,backgrounds,positioning,fit,calc,automata}
\usetikzlibrary{decorations.pathreplacing}
\usetikzlibrary{trees}
\usetikzlibrary{shapes}
\usetikzlibrary{chains}
\usetikzlibrary{patterns}
\usetikzlibrary{graphs}

\newcommand{\fix}[1]{#1}
\newmdenv[
topline=false,
bottomline=false,
rightline=false,
skipabove=\topsep,
skipbelow=\topsep
]{reviewer}

\renewcommand{\paragraph}[1]{\smallskip\noindent\textbf{#1.}}

\newcommand{\cG}{\mathcal{G}}
\newcommand{\cN}{\mathcal{VG}}

\newcommand{\sg}{\subseteq}

\newcommand{\Asg}{\mathcal{A}}

\newcommand{\maxcen}{\textsc{Max}_C}

\newcommand{\mdom}{\mathbb{M}}
\newcommand{\mop}{*}
\newcommand{\mone}{\mathbb{1}}
\newcommand{\bbR}{\mathbb{R}}

\newcommand{\cpf}{f_{\mop, \ell}}

\newcommand{\cpfran}{\operatorname{Range}_{\mop, \ell}}

\tikzset{font={\fontsize{7pt}{12}\selectfont},
	ble/.style={
		circle,
		inner sep=0pt,
		align=center,
		draw=black
	},
	bnode/.style={
		circle,
		inner sep=0pt,
		align=center,
		draw=black,
		fill=bulletcolor!50
	},
	bsnode/.style={
		circle,
		inner sep=0pt,
		align=center,
		draw=black,
		fill=bulletcolor!80
	},
	bgnode/.style={
		circle,
		inner sep=0pt,
		align=center,
		draw=black,
		fill=titlecolor!60
	}
}

\tikzset{
	treenode/.style = {align=center, inner sep=0pt, text centered,
		font=\sffamily},
	arn_n/.style = {treenode, circle, black, font=\sffamily, draw=black,
		fill=white, text width=1.5em},
	arn_r/.style = {treenode, circle, red, draw=red, 
		text width=1.5em, very thick},
	arn_x/.style = {treenode, rectangle, draw=black,
		minimum width=0.5em, minimum height=0.5em}
}

\tikzset{font={\fontsize{7pt}{12}\selectfont},
	ble/.style={
		circle,
		inner sep=0pt,
		align=center,
		draw=black,
	},
	wht/.style={
		circle,
		inner sep=0pt,
		align=center
	},
	snode/.style={
		circle,
		inner sep=0pt,
		align=center,
		draw=black,
		scale=0.15,
		fill=black
	},
	stnode/.style={
		circle,
		inner sep=0pt,
		align=center,
		draw=black,
		scale=0.2
	},
	stextnode/.style={
		circle,
		inner sep=-4pt,
		outer sep=0pt,
		draw=black,
		align=center,
		scale=1
	},
}

\AtBeginDocument{%
	\providecommand\BibTeX{{%
			\normalfont B\kern-0.5em{\scshape i\kern-0.25em b}\kern-0.8em\TeX}}}

\author{Cristian Riveros and Jorge Salas}{Pontificia Universidad Católica de Chile, Chile}{}{}{}

\author{Oskar Skibski}{University of Warsaw, Poland}{}{}{}

\authorrunning{C. Riveros, J. Salas, O. Skibski.} 

\Copyright{Jane Open Access and Joan R. Public} 



\category{} 

\relatedversion{} 

\nolinenumbers 

\begin{document}
\title{\fix{How do centrality measures choose the root of trees?\footnote{\fix{This paper is a resubmission of the article titled ``On the foundations of data centrality:the trees case''. We marked in blue the changes made concerning the previous version. In the appendix, we attached a response letter addressing the main comments provided by the reviewers.}}}}

\maketitle


\begin{abstract}	
\fix{Centrality measures are widely used to assign importance to graph-structured data. Recently, understanding the principles of such measures has attracted a lot of attention. Given that measures are diverse, this research has usually focused on classes of centrality measures. In this work, we provide a different approach by focusing on classes of graphs instead of classes of measures to understand the underlying principles among various measures. More precisely, we study the class of trees. We observe that even in \fix{the} case of trees, there is no consensus on which node should be selected as the most central. 
To analyze the behavior of centrality measures on trees, we introduce a property of \emph{tree rooting} that states a measure selects one or two adjacent nodes as the most important, and the importance decreases from them in all directions. This property is satisfied by closeness centrality but violated by PageRank. We show that, for several centrality measures that root trees, the comparison of adjacent nodes can be inferred by \emph{potential functions} that assess the quality of trees. We use these functions to give fundamental insights on rooting and derive a characterization explaining why some measure root trees. Moreover, we provide an almost liner-time algorithm to compute the root of a graph by using potential functions. Finally, using a family of potential functions, we show that many ways of tree rooting exist with desirable properties.}

\end{abstract}

\section{Introduction}\label{Sec:introduction}


Centrality measures are fundamental tools for network analysis. They are used in a plethora of applications from various areas of science, such as finding people who are more likely to spread a disease in the event of an epidemic~\cite{Dezso:Barabasi:2002} or highlighting cancer genes in proteomic data~\cite{Ivan:Grolmusz:2010}. In all these applications, people use centrality measures to assess graph data and rank which nodes are the most relevant by only using the graph's structure.

\fix{Database systems have also incorporated centrality measures to rank nodes in graph data. The first significant use of a centrality measure in a data management problem is by the Google search engine, which successfully introduced PageRank~\cite{page1999pagerank} for assessing the importance of a website on the web. Indeed, today graph databases, like Neo4j~\cite{neo4j} or TigerGraph~\cite{tigergraph}, natively support centrality algorithms for ranking nodes or analyzing the graph structure. Furthermore, new applications for centrality measures have emerged over knowledge graphs for entity linking~\cite{martinez2020information} or semantic web search engines where ranking results is a core task~\cite{hogan2011searching}. 
More generally, centrality measures play a central role in \emph{network science}, where they are one of the main algorithmic metrics for analyzing graphs~\cite{newman2018networks}.}

\fix{Although more than 200 measures have been proposed in the literature today, it is less clear how one can compare them. Precisely, these measures assess the position of a vertex in the network based on various elements (e.g., degree, number of paths, eigenvector), which makes it hard to compare their properties. Since people propose dozens of new measures every year, choosing a measure for a specific application becomes more and more challenging. In recent years, efforts at understanding and explaining differences between existing measures have intensified, making the foundational aspects of centrality measures an active research topic~\cite{Boldi:etal:2017,Skibski:2021:vitality,RiverosS20,Skibski:etal:2018:gtc}. Given the diversity of these measures, people have usually followed an axiomatic approach by considering classes of measures like game-theoretical~\cite{Skibski:etal:2018:gtc} or motif-based~\cite{RiverosS20} measures. Until now, no research has focused on graph classes to check some similarities between different centrality measures.}

\fix{This paper aims to understand how centrality measures operate on trees (i.e., acyclic undirected graphs), as this is a crucial graph family where we can compare different centrality measures and understand their underlying principles. Indeed, even in the case of trees, centrality measures vary significantly as different measures could indicate other vertices as the most central in the same tree. For example, consider a line graph: while the middle vertex (or vertices) are ranked first according to most centrality measures, Google's PageRank puts the second and the second-to-last vertices at the top of the ranking. This simple example shows that not only do different measures select different vertices, but even one measure may select several vertices from different parts of the tree as the most important. Moreover, although PageRank diverges, other centrality measures (e.g., closeness or all-subgraph) coincide, probably following the same underlying principles.} 

\fix{A natural question here is why we should start considering centrality measures over the class of trees. We see several reasons for studying the principle aspects of centrality measures over trees. First, undirected trees are arguably the simplest and non-trivial graph class to study centrality measures. Indeed, trees have a more amenable structure than general graphs, given that, among other properties, every edge is a cutting edge, and there is a unique path between any two nodes. Second, every general result on centrality measures should include trees, and it should answer similar questions to the one studied in this paper. Thus, understanding centrality measures over trees could guide the study of other graph families. Third and last, trees are probably the most ubiquitous graph structure in computer science and databases. Therefore, studying centrality measures over trees could be helpful for the application of centrality measures in data management and other areas (see Section~\ref{Sec:concl} for some extensions and possible applications of centrality measures over trees).}

\fix{What principle aspect of centrality measures can we study over the class of trees? In this paper, we focus on understanding one of the main questions over trees: how centrality measures choose the most central nodes in trees, why some measures define a single important node (usually called the root), and why others do not. } For answering these questions, the main contributions are as follows:
\begin{enumerate}
	\item We introduce the \emph{tree rooting property} which states \fix{that} a measure selects one or two adjacent vertices in a tree as the most important, and the importance decreases from them in all directions. We found that closeness, eccentricity, and all-subgraphs centralities (see Section 2) satisfy this rooting property but often rank different vertices at the top. Instead, measures like degree and betweenness do not satisfy this property. We call the vertex (or two vertices) ranked first the ``root'' and say \fix{that} such measures \emph{root~trees}.
	
	\item To understand what distinguishes measures that root trees we focus on the question: how to choose which out of two adjacent vertices is more central? 
	We observe that most centrality measures, including all standard ones that root trees, answer this question by comparing subtrees of both vertices.
	More precisely, we introduce a framework of \emph{potential functions} that assess the quality or ``potential'' of an arbitrary tree.
	Now, we show \fix{that} most centrality measures admit a potential function such that the vertex which subtree has higher potential value is considered more central.
	
	\item We show that if a centrality measure has a potential function, then it roots trees if, and only if, the potential function is \emph{symmetric}.
	This property means that the potential of a tree is larger than the potential of any proper complete subtree. 
	In particular, the potential function of closeness, eccentricity, and all-subgraphs centrality are symmetric, but the potential functions of degree and betweenness centralities are not; as a result, the later centralities do not root trees.
	
	\item We use our framework of potential functions to understand better the class of measures that root trees. More specifically, we show three applications of potential functions. Our first application is to study efficient algorithms for computing a root when centrality measures have potential functions and root trees. 
	By exploiting symmetric potential functions, we show that, given a tree $T$, we can compute the most central vertex in time $\mathcal{O}(|T|\log(|T|))$ whenever one can calculate the potential function locally. Interestingly, this general algorithm works independently of the centrality measures and only depends on the potential function.
	
	\item Our second application of potential functions is to understand desirable properties over tree rooting measures. Although a centrality measure could root trees, it can behave inconsistently. For instance, a rooting centrality measure could choose the root of a tree $T$ close to a leaf when the size of $T$ is even and far from the leaves when the size of $T$ is odd. Therefore, we study when centrality measures consistently root trees through their potential functions. We propose a monotonicity property that imposes additional consistency conditions on how the root is selected and characterizes which potential functions satisfy this property. 
	
	\item Our last application shows how to design and build new potential functions that consistently root trees. Specifically, we present infinitely many constructive potential functions that satisfy all properties discussed so far. In particular, the algorithm for finding the root applies to any of them. We believe that this family of measures is interesting in its own right and can be used in several data-driven scenarios.
\end{enumerate}


\paragraph{Related work}
Our work could be included into a broad literature that focuses on the analysis of theoretical properties of centrality measures.
The classic approach is to analyze standard centrality measures with respect to simple desirable properties.
Different properties have been considered over the years \cite{sabidussi1966centrality,Nieminen:1973,boldivigna,Boldi:etal:2017}.
\fix{Most of them, however, are very simple (e.g., invariance under automorphism) or not satisfied by most measures.
Similarly, in our work, we propose several properties specific for trees. 
Focusing on trees allows us to identify meaningful properties shared by many measures based on completely different principles.}

Another approach is to create a common framework for large classes of centrality measures. 
In such frameworks, centrality measures are presented as a function of some underlying structure of the graph.
Hence, the emphasis is focused on the differences between these functions and their implications for various measures defined under such framework.
\fix{In this spirit, classes of measures based on distances~\cite{garg2009axiomatic}, nodal statistics~\cite{Bloch:etal:2019}, coalitional game theory~\cite{Skibski:etal:2018:gtc}, subgraphs~\cite{RiverosS20} and vitality functions~\cite{Skibski:2021:vitality} have been analyzed in the literature.
On the opposite, our framework of potential functions can be considered as an approach focused on classes of graphs instead of classes of measures.}
Yet another approach, less related to our work, is to focus on one or several similar measures and provide their full axiomatic characterization. 
In this spirit, axiomatizations of PageRank~\cite{Was:Skibski:2018:pagerank}, eigenvector \cite{Kitti:2016}, beta-measure and degree \cite{Brink:Gilles:2000} and many more have been developed in the literature.

There is also a line of research that studies methods for solely choosing one most central vertex in a tree that does not necessarily come from the centrality analysis (see, e.g., \cite{Reid:2011} for an overview).
However, such methods coincide with the top vertices selected by centrality measures that we consider in our work. In particular, the center of a tree coincides with eccentricity, and the median, as well as the centroid, coincides with closeness centrality.


\section{Preliminaries}\label{Sec:preliminaries}


\paragraph{Undirected graphs} In this paper, we consider \emph{finite undirected graphs} of the form $G = (V, E)$ where $V$ is a finite non-empty set and $E \subseteq 2^{V}$ such that $|e|=2$ for all $e \in E$. For convenience, given a graph $G=(V,E)$ we use $V(G)$ for indicating the set of \emph{vertices} $V$ and $E(G)$ for the set of \emph{edges} $E$.
We write $N_G(v)$ for the \emph{neighbourhood} of $v$ in $G$, namely, $N_G(v) \subseteq V(G)$ such that $u \in N_G(v)$ if, and only if, $\{u,v\} \in E(G)$. If this is the case, we say that $u$ and $v$ are \emph{adjacent}.

We say that a graph $G' = (V', E')$ is a \emph{subgraph} of $G$, denoted $G' \sg G$, if $V' \subseteq V$ and $E' \subseteq E$.  Note that $\sg$ forms a partial order between graphs. 
For a sequence of graphs $G_1 = (V_1, E_1), \ldots, G_m = (V_m, E_m)$ we denote by $\cup_{i=1}^m G_i$ the graph $(V, E)$ such that $V = \bigcup_{i=1}^m V_i$ and $E = \bigcup_{i=1}^m E_i$. Given a graph $G$ and an edge $e = \{u,v\}$, we write $G + e$ to be the new graph $G$ with the additional edge $e$, formally, $V(G+e) = V(G) \cup e$ and $E(G+e) = E(G) \cup \{e\}$. 

From now on, fix \fix{an enumerable} set $\mathcal{V}$ of vertices. We define the \emph{set of all graphs} using vertices from $\mathcal{V}$ as~$\cG$. Further, we define the set $\cN$ as all pairs \emph{vertex-graph} $(v, G)$ such that $v \in V(G)$ and $G \in \cG$.  In the sequel, for a pair $(v, G)$ we assume that $(v, G) \in \cN$, unless stated otherwise. We say that graphs $G_1$ and $G_2$ are \emph{isomorphic}, denoted by $G_1 \cong G_2$, if there exists an bijective function (\emph{isomorphism}) $f:V(G_1) \rightarrow V(G_2)$ such that $\{u,v\} \in E(G_1)$ if, and only if, $\{f(u), f(v)\} \in E(G_2)$. We also say that $(v_1, G_1)$ and $(v_2, G_2)$ are isomorphic, denoted by $(v_1, G_1) \cong (v_2, G_2)$, if there exists an isomorphism $f$ between $G_1$ and $G_2$ and $f(v_1) = v_2$. Note that $\cong$ is an equivalence relation over $\cG$ and over $\cN$.

A \fix{\emph{path}} in $G$ is a sequence of vertices $\pi = v_0, \ldots, v_n$ such that $\{v_i, v_{i+1}\} \in E$ for every $i < n$, and we say that $\pi$ is a \emph{path} from $v_0$ to $v_n$. We say that $\pi$ is \emph{simple} if $v_i \neq v_j$ for every $0 \leq i < j < n$. 
From now on, we usually assume that paths are simple unless stated otherwise.
We define the \emph{length} of $\pi$ as $|\pi| = n$. 
We agree that $v_0$ is the \emph{trivial path} of length $0$ from $v_0$ to itself. 
Given $R, R' \subseteq V(G)$, we say that $\pi = v_0, \ldots, v_n$ is a path from $R$ to $R'$ if $v_0 \in R$, $v_n \in R'$, and $v_i \notin R \cup R'$ for every $i \in [1, n-1]$. 
We say that a graph $G$ is \emph{connected} if there exists a path between \fix{every} pair of vertices.



\paragraph{Centrality measures} A \emph{centrality measure}, or just a \emph{measure}, is any function $C: \cN \to\mathbb{R}$ that assigns a score $C(v, G)$ to $v$ depending on its graph $G$. 
Here, we use the standard assumption that, the higher the score $C(v, G)$, the more important or ``central'' is $v$ in~$G$. 
We also assume that every centrality measure is \emph{closed under isomorphism}, namely, if $(v_1, G_1) \cong (v_2, G_2)$ then $C(v_1, G_1) = C(v_2, G_2)$, which is a standard assumption in the literature \cite{brandes2005network,sabidussi1966centrality}.

Next, we recall four centrality measures that we will regularly use as examples: \emph{degree}, \emph{closeness}, \emph{eccentricity}, and \emph{all-subgraphs} centralities. During this work, we also mention \emph{betweenness}~\cite{freeman1977set}, \emph{decay}~\cite{jackson2010social}, \emph{PageRank}~\cite{page1999pagerank}, and \emph{eigenvector}~\cite{bonacich1972factoring} centralities. Given that we do not use them directly, their definitions are in the appendix.

\paragraph{Degree centrality} Degree centrality is probably the most straightforward measure. Basically, the bigger the neighborhood of a vertex (i.e., adjacent vertices), the more central it is in the graph. Formally, we define \emph{degree centrality} as follows:
$
\textsc{Degree}(v,G) = |N_G(v)|
$.

\paragraph{Closeness centrality} For every graph $G$ and vertices $v,u \in V(G)$ we define the \emph{distance} between $v$ and $u$ in $G$ as $d_G(v,u)= |\pi_{v,u}|$ where $\pi_{v,u}$ is a shortest path from $v$ to $u$ in $G$. 
Then \emph{closeness centrality}~\cite{sabidussi1966centrality} is defined as: $\textsc{Closeness}(v,G) = 1/\sum_{u\in K_v(G)} d_{G}(v,u)$
where $K_v(G)$ is the connected component of $G$ containing $v$.
Closeness is usually called a geometrical measure because it is based on the distance inside a graph. 
The intuition behind closeness centrality is simple: the closer a vertex is to everyone in the component (i.e., $\sum_{u\in K_v(G)} d_{G}(v,u)$ is small) the more important it is.

\paragraph{Eccentricity centrality} Another important notion in graph theory is radius. In simple words, we can define the center of a graph $G$ as the vertex that \fix{minimizes} the maximum distance in $G$. Formally, $v$ is the center of $G$ if it minimizes 
\fix{$\max_{u\not=v\in V(G)}d_{G}(v,u)$}. Then, the radius of $G$ is defined as the \fix{maximum distance from the center}. Now, \emph{eccentricity measure}~\cite{hage1995eccentricity} is precisely the one centrality that \fix{selects} the center of a graph as the most important vertex, defined as
$\textsc{Eccentricity}(v,G) = 1/\max_{u\in V(G)}d_{G}(v,u)$.


\paragraph{All-subgraphs centrality}  Given a graph $G = (V, E)$ and a vertex $v \in V$, we denote by $\Asg(v,G)$ the set of all connected subgraphs of $G$ that \fix{contain} $v$, formally, $\Asg(v,G) = \{S \subseteq G \mid v \in V(S) \text{ and } S \text{ is connected}\}$. 
Then \emph{all-subgraphs centrality}~\cite{RiverosS20} of $v$ in $G$ is defined as: 
$
\textsc{AllSubgraphs}(v,G) = \log_2|\Asg(v,G)|
$.
All-subgraphs centrality was recently proposed in~\cite{RiverosS20}, proving that it satisfies several desirable properties as a centrality measure. Intuitively, it says that a vertex will be more relevant in a graph if it has more connected subgraphs surrounding it.  

\paragraph{Undirected Trees} This paper is about \emph{undirected trees} (or just trees), so we use some special notation for them.
Specifically, we say that a graph $T$ is a \emph{tree} if it is connected and for every $u,v \in V(T)$ there exists a unique path that connects $u$ with $v$ in $T$. We usually use $T$ to denote a tree. Further, we say that $v \in V(T)$ is a \emph{leaf} of $T$ if $|N_T(v)| = 1$. If $v$ is a leaf and $N_T(v) = \{u\}$, then we say that $u$ is the \emph{parent} of $v$. Note that trees are a special class of undirected graphs, and \fix{all previous} definitions apply. In particular, we can use and apply centrality measures over trees. 

We say that $T'$ is a \emph{subtree} of $T$ if $T' \subseteq T$ and $T'$ is a tree. We also say that $T'$ is a \emph{complete subtree} of $T$ if $T' \subseteq T$ and there is at most one vertex in $V(T')$ connected to some vertex in $V(T) \setminus V(T')$, namely, $|\{v \in V(T') \mid \exists u \in V(T) \setminus V(T').\,  \{u, v\} \in E(T)\}| \leq 1$.

The following notation will be useful in the paper to decompose trees.
Given a tree $T$ and two adjacent vertices $u, v \in V(T)$, we denote by $T_{u,v}$ the maximum subtree of $T$ that contains $u$ and not $v$. For example, if $T = $ \raisebox{-0.8mm}{\mbox{
\begin{tikzpicture}[node distance=2mm,text width=5mm]
\node (u) at (0,0) [stnode] {};

\draw [-, line width=0.2pt, rounded corners=0.8] (u) -- (-0.4,0.15) -- (-0.4,-0.15) -- (u);

\node (u') at (0.3,0) [stnode, fill=black] {};

\draw [-, line width=0.2pt, rounded corners=0.8] (u') -- (0.7,0.15) -- (0.7,-0.15) -- (u');

\draw [-, line width=0.1pt] (u) to (u');

\end{tikzpicture}}}  
, then $T_{\circ, \bullet} = \raisebox{-0.8mm}{\mbox{
\begin{tikzpicture}[node distance=2mm,text width=5mm]
\node (u) at (0,0) [stnode] {};

\draw [-, line width=0.2pt, rounded corners=0.8] (u) -- (-0.4,0.15) -- (-0.4,-0.15) -- (u);

%
%

\end{tikzpicture}}}  
$  \,\! and \,\! $T_{\bullet, \circ} = \raisebox{-0.8mm}{\mbox{
\begin{tikzpicture}[node distance=2mm,text width=5mm]
%

\node (u') at (0.3,0) [stnode, fill=black] {};

\draw [-, line width=0.2pt, rounded corners=0.8] (u') -- (0.7,0.15) -- (0.7,-0.15) -- (u');


\end{tikzpicture}}}  
$ \ .

Finally, we consider some special trees to give examples or show some properties of centrality measures. 
For a vertex $v$ we define  $G_v=(\{v\}, \emptyset)$, and for an edge $e$ we define $G_e = (e, \{e\})$, namely, the graphs with \emph{one isolated vertex} $v$ or \emph{one isolated edge} $e$, respectively. 
Similarly, for any $n \geq 1$ we write $L_n$ for the \emph{line} with $n$ vertices where $V(L_n) = \{0, \ldots, n-1\}$ and $E(L_n) = \{\{i, i+1\} \mid 0 \leq i < n-1\}$.

\section{Tree rooting centrality measures}\label{Sec:rooting}
We start by giving a formal definition of when a centrality measure $C$ roots trees. For this, let $C$ be a centrality measure. We define the set $\maxcen(T)$ to be the set of most central vertices with respect to $C$ in a tree $T$, namely, $v \in \maxcen(T)$ iff $C(u, T) \leq  C(v, T)$ for every~$u \in V(T)$.

 
\begin{definition}\label{def:rooting}
	We say that a centrality measure $C$ \emph{roots trees} if for every tree $T$, the set of most central vertices $\maxcen(T)$ consists of one vertex or two adjacent vertices. Moreover, for \fix{every} $u \notin \maxcen(T)$ if $u_0 u_1\fix{\ldots} u_n$ is the unique path from $\maxcen(T)$ to $u$, then $C(u_{i}, T) > C(u_{i+1}, T)$ for every $i\in [0,n-1]$.
\end{definition}


In the following, if a centrality measure roots trees, we also say that it satisfies the \emph{tree rooting property} (i.e., Definition~\ref{def:rooting}). 
We can motivate this property as follows. 
We treat vertices with the highest centrality as ``roots''.
For the first part of the definition, we assume there is a single source of importance -- one vertex or two adjacent vertices.
We allow two adjacent vertices to be the roots, as in some graphs, due to their symmetrical structure, it is impossible to indicate one most central vertex. 
This is, for example, the case of a line \fix{with an even} number of vertices (e.g.,
{\mbox{
\begin{tikzpicture}[node distance=2mm,text width=5mm]
\node (u1) at (0,0) [stnode] {};
\node (u2) at (0.3,0) [stnode] {};
\node (u3) at (0.6,0) [stnode] {};
\node (u4) at (0.9,0) [stnode] {};

\draw [-, line width=0.1pt] (u1) to (u2);
\draw [-, line width=0.1pt] (u2) to (u3);
\draw [-, line width=0.1pt] (u3) to (u4);

\end{tikzpicture}}}  
\!).
In such a scenario the edge between both can be considered the real root of the tree.
For the second part, we assume that the centrality should decrease from the root through branches. This restriction aligns with the intuition that the closer a vertex is to the root, the more central it is.


We continue by giving examples of measures that root trees, and some others that do not.

\newcommand{\lmspace}{\hfill}
\begin{figure*}
	\lmspace
	\begin{minipage}{0.2\textwidth}
		\begin{tikzpicture}[node distance=0.9cm] 
	\tikzset{     
		e4c node/.style={circle,draw,minimum size=0.35cm,inner sep=0}, 
		e4c edge/.style={sloped,above}
	}
	
	\node[e4c node, fill=black!56] (5) at (0, 0) {}; 
	
	\node at (0, 1) {}; 
	
	\node[e4c node, right of=5, fill=black!28] (1) {}; 
	\node[e4c node, above right of=5, fill=black!28] (2) {}; 
	\node[e4c node, above left of=5, fill=black!28] (3)  {}; 
	\node[e4c node, left of=5, fill=black!28] (4) {}; 
	
	\node[e4c node, below of=5,minimum size=0.5cm, fill=black!70, star] (6) {}; 
	\node[e4c node, below of=6, fill=black!56] (7) {}; 
	\node[e4c node, left of=7, fill=black!42] (8) {}; 
	\node[e4c node, below of=8, fill=black!14] (9) {}; 
	\node[e4c node, right of=9, fill=black!1] (10) {}; 
	\node[e4c node, right of=7, fill=black!28] (11) {};

	\path[draw,thick]
	(1) edge[e4c edge]  (5)
	(4) edge[e4c edge]  (5)
	(2) edge[e4c edge]  (5)
	(3) edge[e4c edge]  (5)
	(5) edge[e4c edge]  (6)
	(6) edge[e4c edge]  (7)
	(7) edge[e4c edge]  (11)
	(7) edge[e4c edge]  (8)
	(8) edge[e4c edge]  (9)
	(9) edge[e4c edge]  (10)
	;
	
	\node at (0, -3.3) {\textsc{\large Closeness}};
\end{tikzpicture}
	\end{minipage} \lmspace
	\begin{minipage}{0.2\textwidth}
		\begin{tikzpicture}[node distance=0.9cm] 
	\tikzset{     
		e4c node/.style={circle,draw,minimum size=0.35cm,inner sep=0}, 
		e4c edge/.style={sloped,above}
	}
	
	\node[e4c node, fill=black!24] (5) at (0, 0) {}; 
	
	\node at (0, 1) {}; 
	
	\node[e4c node, right of=5, fill=black!1] (1) {}; 
	\node[e4c node, above right of=5, fill=black!1] (2) {}; 
	\node[e4c node, above left of=5, fill=black!1] (3)  {}; 
	\node[e4c node, left of=5, fill=black!1] (4) {}; 
	
	\node[e4c node, below of=5, fill=black!47] (6) {}; 
	\node[e4c node, below of=6, fill=black!70, minimum size=0.5cm, star] (7) {}; 
	\node[e4c node, left of=7, fill=black!47] (8) {}; 
	\node[e4c node, below of=8, fill=black!24] (9) {}; 
	\node[e4c node, right of=9, fill=black!1] (10) {}; 
	\node[e4c node, right of=7, fill=black!47] (11) {}; 
	
	ss
	\path[draw,thick]
	(1) edge[e4c edge]  (5)
	(4) edge[e4c edge]  (5)
	(2) edge[e4c edge]  (5)
	(3) edge[e4c edge]  (5)
	(5) edge[e4c edge]  (6)
	(6) edge[e4c edge]  (7)
	(7) edge[e4c edge]  (11)
	(7) edge[e4c edge]  (8)
	(8) edge[e4c edge]  (9)
	(9) edge[e4c edge]  (10)
	;
	
	\node at (0, -3.3) {\textsc{\large Eccentricity}};
\end{tikzpicture}
	\end{minipage} \lmspace
	\begin{minipage}{0.2\textwidth}
		\begin{tikzpicture}[node distance=0.9cm] 
	\tikzset{     
		e4c node/.style={circle,draw,minimum size=0.35cm,inner sep=0}, 
		e4c edge/.style={sloped,above}
	}
	
	\node[e4c node, fill=black!70, minimum size=0.5cm, star] (5) at (0, 0) {}; 
	
	\node at (0, 1.5) {}; 
	
	\node[e4c node, right of=5, fill=black!30] (1) {}; 
	\node[e4c node, above right of=5, fill=black!30] (2) {}; 
	\node[e4c node, above left of=5, fill=black!30] (3)  {}; 
	\node[e4c node, left of=5, fill=black!30] (4) {}; 
	
	\node[e4c node, below of=5, fill=black!60] (6) {}; 
	\node[e4c node, below of=6, fill=black!50] (7) {}; 
	\node[e4c node, left of=7, fill=black!40] (8) {}; 
	\node[e4c node, below of=8, fill=black!20] (9) {}; 
	\node[e4c node, right of=9, fill=black!1] (10) {}; 
	\node[e4c node, right of=7, fill=black!10] (11) {};

	\path[draw,thick]
	(1) edge[e4c edge]  (5)
	(4) edge[e4c edge]  (5)
	(2) edge[e4c edge]  (5)
	(3) edge[e4c edge]  (5)
	(5) edge[e4c edge]  (6)
	(6) edge[e4c edge]  (7)
	(7) edge[e4c edge]  (11)
	(7) edge[e4c edge]  (8)
	(8) edge[e4c edge]  (9)
	(9) edge[e4c edge]  (10)
	;
	
	\node at (0, -3.3) {\textsc{\large AllSubgraphs}};
\end{tikzpicture}
	\end{minipage} \lmspace
	\begin{minipage}{0.2\textwidth}
		\begin{tikzpicture}[node distance=0.9cm] 
	\tikzset{     
		e4c node/.style={circle,draw,minimum size=0.35cm,inner sep=0}, 
		e4c edge/.style={sloped,above}
	}
	
	\node[e4c node, fill=black!70, minimum size=0.5cm, star] (5) at (0, 0) {}; 
	
	\node at (0, 1) {}; 
	
	\node[e4c node, right of=5, fill=black!1] (1) {}; 
	\node[e4c node, above right of=5, fill=black!1] (2) {}; 
	\node[e4c node, above left of=5, fill=black!1] (3)  {}; 
	\node[e4c node, left of=5, fill=black!1] (4) {}; 
	
	\node[e4c node, below of=5, fill=black!35] (6) {}; 
	\node[e4c node, below of=6, fill=black!70, minimum size=0.5cm, star] (7) {}; 
	\node[e4c node, left of=7, fill=black!1] (8) {}; 
	\node[e4c node, below left of=7, fill=black!1] (9) {}; 
	\node[e4c node, below right of=7, fill=black!1] (10) {}; 
	\node[e4c node, right of=7, fill=black!1] (11) {};

	\path[draw,thick]
	(1) edge[e4c edge]  (5)
	(4) edge[e4c edge]  (5)
	(2) edge[e4c edge]  (5)
	(3) edge[e4c edge]  (5)
	(5) edge[e4c edge]  (6)
	(6) edge[e4c edge]  (7)
	(7) edge[e4c edge]  (11)
	(7) edge[e4c edge]  (8)
	(7) edge[e4c edge]  (9)
	(7) edge[e4c edge]  (10)
	;
	
	\node at (0, -3.3) {\textsc{\large Others}};
\end{tikzpicture}
	\end{minipage} 
	\lmspace
	
\vspace{-3mm}	
	\caption{The first three trees exemplify how closeness, eccentricity, and all-subgraphs centralities root trees. We mark the most central vertex with a black star. The colors show how the centrality value decreases through the branches (i.e., whiter vertices are less central). The fourth tree is a counter-example that shows why other centralities do not root trees.}
	\label{fig:rooting-examples}
\end{figure*}
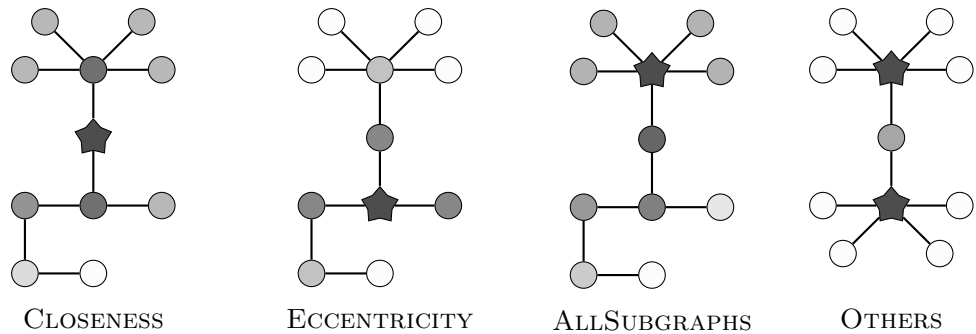
\begin{example}
Closeness, eccentricity, and all-subgraphs all root trees. It was already noticed~\cite{koschutzki2005centrality} that over trees, closeness and eccentricity define at most two maximum vertices, and both are connected. \fix{On the other hand}, it is more subtle to show that all-subgraphs centrality \fix{roots} trees. 
This fact, however, will follow from the framework developed in Section~\ref{Sec:potential}.
As an illustration, in Figure~\ref{fig:rooting-examples} we show how closeness, eccentricity, and all-subgraphs behave over the same tree. One can verify that each measure declares one vertex with the maximum centrality (this vertex is marked with a black star).
Moreover, the centrality decreases through the branches (the lower the centrality the whiter the color). 
It is interesting that, although the three measures root trees, they declare different vertices as the most central.
\end{example}



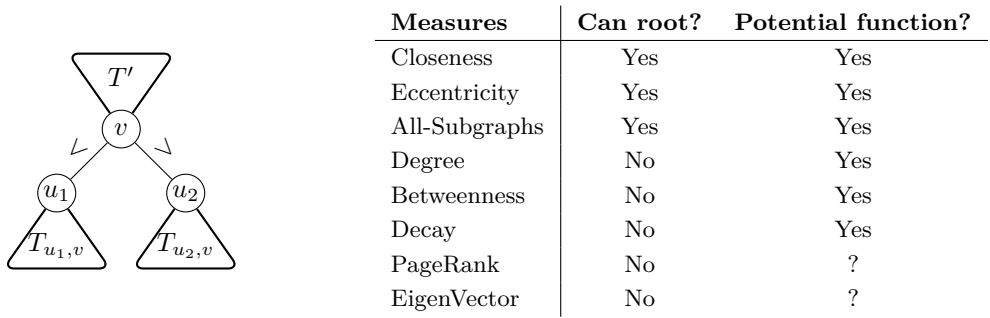
\begin{figure}[t]
	\hfill
	\begin{minipage}{0.3\textwidth}
		\newcommand{\fsize}{\normalsize}
\begin{tikzpicture}[node distance=1.2cm] 
	\tikzset{     
		e4c node/.style={circle,draw,minimum size=0.5cm,inner sep=0}, 
		e4c edge/.style={sloped,above}
	}
	
	\node[e4c node] (v) at (0, 0) {\fsize $v$}; 
	
	\node[e4c node, below left of=v] (u1) {\fsize $u_1$}; 
	\node[e4c node, below right of=v] (u2) {\fsize $u_2$};

	\path[draw]
	(v) edge[e4c edge] node {\fsize $<$}  (u1)
	(v) edge[e4c edge] node {\fsize $>$}  (u2)
	;
	
	\draw [draw, thick, rounded corners] (v) -- ($(v) + (-0.7,1)$) -- ($(v) + (0.7,1)$) -- (v);
	
	\draw [draw, thick, rounded corners] (u1) -- ($(u1) + (-0.7,-1)$) -- ($(u1) + (0.7,-1)$) -- (u1);
	
	\draw [draw, thick, rounded corners] (u2) -- ($(u2) + (-0.7,-1)$) -- ($(u2) + (0.7,-1)$) -- (u2);
	
	\node[node distance=0.7cm, above of=v] {\fsize $T'$}; 
	\node[node distance=0.7cm, below of=u1] {\fsize $T_{u_1,v}$};
	\node[node distance=0.7cm, below of=u2] { \fsize $T_{u_2,v}$}; 
\end{tikzpicture}
	\end{minipage}  
\hfill
	\begin{minipage}{0.6\textwidth}
		\begin{tabular}{l|c c}
			\textbf{Measures}         & \textbf{Can root?} &  \textbf{Potential function?} \\ \hline
			Closeness       & Yes & Yes      \\
			Eccentricity    & Yes & Yes      \\
			All-Subgraphs    & Yes & Yes      \\
			Degree      & No & Yes        \\
			Betweenness      & No & Yes       \\
			Decay           & No  & Yes      \\
			PageRank        & No  & ?      \\
			EigenVector     & No  & ?      
		\end{tabular}
	\end{minipage} 
\hfill

	\caption{At the left, a graphic illustration of the symmetry property over a tree $T$. Subtrees $T_{u_1, v}$ and $T_{u_2, v}$ are isomorphic, and subtree $T'$ represents the rest of $T$ hanging from~$v$. At the right, a table summarizes which centrality measures root trees and which one admits a potential function.}
	\label{fig:symmetry}
\end{figure}

\begin{example}
	One can easily check that degree centrality does not root trees. Indeed, the last tree at Figure~\ref{fig:rooting-examples} is an example where degree centrality declares two maximum vertices, and they are not adjacent. Indeed, all measures presented in Section~\ref{Sec:preliminaries}, except closeness, eccentricity, and all-subgraphs, do not root trees. For all of them, the last tree at Figure~\ref{fig:rooting-examples} is a counterexample where they violate the tree rooting property. At Figure~\ref{fig:symmetry}, we show a table that summarize which centrality measures considered in this paper root trees.
\end{example}

An important consequence of assigning a root to a tree is that each vertex has a parent (except for the root). Here, the unique path from the root to the vertex defines its parent. Another possibility would be to use the centrality measure to find the neighbour with higher centrality, and declare it as the parent. We capture \fix{this intuition} in the following property. 


\begin{definition}
	We say that a measure $C$ satisfies the \emph{at-most-one-parent} property if for every tree $T$ and $v \in V(T)$ there exists at most one neighbour $u \in N_T(v)$ such that~$C(v,T) \leq~C(u, T)$. 
\end{definition}

Clearly, if a centrality measure $C$ roots trees, then it also satisfies at-most-one-parent. The other direction is also true, providing an alternative characterization for tree rooting. 

\begin{proposition}\label{prop:amoproot}
A centrality measure roots trees if, and only if, it satisfies the at-most-one-parent property. 
\end{proposition}

Notice that tree rooting is a global property over a tree but, instead, at-most-one-parent is a local property of the neighbourhoods of a tree, which is easier to prove for a centrality measure. Indeed, in the next section we use this alternative definition to prove our main characterization for tree rooting.

In some trees the root is uniquely characterized solely by the tree rooting property.
This happens because every centrality must be closed under isomorphism which \fix{implies that isomorphic} vertices have the same centrality. For example, one can verify that for the line $L_n$ the roots must be the set $\{\lfloor\frac{n-1}{2}\rfloor, \lceil\frac{n-1}{2}\rceil\}$. Indeed, every vertex $i\leq \frac{n-1}{2}$ is isomorphic with the vertex $n-1-i$ in $L_n$. Then, if $i$ is the root, then $n-1-i$ must be the root as well. 
Given that vertices $i$ \fix{and} $n-1-i$ are not connected if $i < \frac{n-1}{2}$, we get that every centrality that roots trees must declare $\{\lfloor\frac{n-1}{2}\rfloor, \lceil\frac{n-1}{2}\rceil\}$ as the root.
This idea can be extended to \fix{all} symmetric trees: if $T$ is a tree with a vertex $\circ$ that connects two isomorphic subtrees (i.e., $T =$ \raisebox{-0.8mm}{\mbox{
\begin{tikzpicture}[node distance=2mm,text width=5mm]
\node (u) at (0,0) [stnode] {};

\draw [-, line width=0.2pt, rounded corners=0.8] (u) -- (-0.2,-0.05) -- (-0.35,-0.3) -- (-0.05,-0.3) -- (-0.2,-0.05);

\draw [-, line width=0.2pt, rounded corners=0.8] (u) -- (0.2,-0.05) -- (0.05,-0.3) -- (0.35,-0.3) -- (0.2,-0.05);




\end{tikzpicture}}}  
), then $\circ$ must be the root of $T$. We generalize this property as follows.

\begin{definition}
	We say that a centrality measure $C$ is \emph{symmetric over trees} if for every tree $T$, vertex $v$, and different neighbors $u_1, u_2 \in N_T(v)$ such that $(u_1, T_{u_1, v}) \cong (u_2, T_{u_2, v})$, then $C(u_1, T) < C(v, T)$ and $C(u_2, T) < C(v, T)$. 
\end{definition}

Figure~\ref{fig:symmetry} (left) is a graphical representation of the symmetry property. 
It generalizes the previously discussed intuition by considering any pair of isomorphic subtrees in a (not necessarily symmetric) tree.
Interestingly, every centrality measure that roots trees must be symmetric over trees. 

\begin{proposition}\label{pr:RootSym}
	If a centrality measure $C$ roots trees, then $C$ is symmetric over trees.
\end{proposition}

A symmetric centrality measure may not root trees. For example, a centrality measure could root trees with non-trivial automorphisms but not root trees when there is none. Despite this, symmetry property is crucial for finding a characterization for tree rooting, as we will show in the next section.

\section{Potential functions}\label{Sec:potential}

\fix{What have in common closeness, eccentricity, and all-subgraphs centralities?} What is the fundamental property so they can root trees? 
A crucial ingredient for understanding the connection between these measures is what we call a \emph{potential function} for a centrality measure. 

\begin{definition}
Given a centrality measure $C$, we say that $f: \cN \to\mathbb{R}$ is a potential function for $C$ if $f$ is closed under isomorphism on $\cN$ and, for every tree $T$ and every adjacent vertices $u, v \in V(T)$, it holds that $C(u, T) \leq C(v,T)$ if, and only if, $f(u, T_{u,v}) \leq f(v, T_{v,u})$.
\end{definition}

A potential function is a function that measures the ``potential'' of \fix{every \emph{rooted} tree, i.e., a tree with one node selected and the assessment depends on the selection.} 
Now, a centrality measure admits some potential function if the comparison between two adjacent vertices is determined by the potential of their corresponding subtrees.
Interestingly, in the following examples we show that several centrality measures admit a potential function. 


\begin{example}\label{ex:pot-func}
Degree, closeness, eccentricity, all-subgraphs centralities have the following potential functions:
	$$
	\begin{array}{rcll}
		f_{\text{d}}(v, T) & := & |N_T(v)| & (degree)\\
		f_{\text{c}}(v, T) & := & |V(T)| & (closeness)\\
		f_{\text{e}}(v, T) & := & \max_{u\in V(T)} d_{T}(v, u) & (eccentricity)\\
		f_{\text{a}}(v, T) & := & |\Asg(v,T)| & (all\text{-}subgraphs)
	\end{array}
	$$
Let us verify that each function above is a potential function for its corresponding measure.
Take an arbitrary tree $T$ and two adjacent vertices $u$ and $v$. 
It is straightforward to check \fix{that $f_{\text{d}}$} is a potential function for degree centrality. 
Indeed, if $u$ has a smaller degree in its subtree (i.e., without considering the common edge), then \fix{it also has} a smaller centrality.

For closeness centrality, the potential function $f_{\text{c}}$ is simply the number of vertices in a tree. 
To see this, note that vertex $u$ has a distance smaller by one than $v$ to all vertices from $T_{u,v}$; analogously, vertex $v$ has a distance smaller by one than $u$ to all vertices from $T_{v,u}$. 
As a result, out of both vertices, the one with the larger subtree has the smaller sum of distances which \fix{results} in the higher closeness~centrality.

For eccentricity the potential function $f_{\text{e}}$ is the height of a tree, i.e., the distance to the farthest vertex. 
This is because if subtree $T_{u,v}$ is higher than $T_{v,u}$, then vertex $u$ has smaller distance to the farthest vertex in $T$ which results in the higher eccentricity. 
Interestingly, $f_{\text{e}}$ is an inverse of eccentricity.

Finally, for all-subgraphs centrality the potential function $f_{\text{a}}$ is the number of subgraphs that contain vertex $v$.
The reason is that we have $|\Asg(u,T)| = |\Asg(u,T_{u,v})| + |\Asg(u,T_{u,v})| \cdot |\Asg(v,T_{v,u})|$ and, symmetrically, $|\Asg(v,T)| = |\Asg(v,T_{v,u})| + |\Asg(u,T_{u,v})| \cdot |\Asg(v,T_{v,u})|$.
This implies that if $u$ has more subgraphs in $T_{u,v}$ than $v$ in $T_{v,u}$, then it also has higher all-subgraphs centrality.
Hence, as in the case of degree centrality, \fix{the potential function} coincides with the centrality itself.
	
%
\end{example}


In the appendix, we also show that betweenness and decay centralities have a potential function.
As we will see later, there are centrality measures that do not have potential functions. For the particular case of PageRank and EigenVector it is not clear whether they admit a potential function. \fix{The table at} Figure~\ref{fig:symmetry} (right) summarizes which centrality measures have a potential function.

A potential function determines the centrality order between two adjacent vertices, but it does not imply the relation between non-adjacent vertices. Although this information is weaker than the centrality measure itself, it is exactly what we need to understand the centrality measures that root trees. Precisely, which measures with potential functions root trees? To answer this question, we first need to capture the symmetry property through the lens of potential functions. 

\begin{lemma}\label{Lemma:SymmetricPot}
	Let $C$ be a centrality measure and $f$ a potential function for $C$. Then $C$ is symmetric over trees if, and only if, for every tree $T$ and every pair of adjacent vertices $u,v \in V(T)$, it holds that $f(u, T_{u,v}) < f(v, T)$. 
\end{lemma}

By the previous result, we will call potential functions with this property \emph{symmetric}. Next, we show that symmetric potential functions characterize the tree rooting property. 

\begin{theorem}\label{theo:rooting}
	Let $C$ be a centrality measure that admits a potential function $f$. Then $C$ roots trees if, and only if, $f$ is symmetric. 
\end{theorem}


%
\begin{example} 
	Continuing with Example~\ref{ex:pot-func} we can check that the potential functions $f_{\text{c}}$, $f_{\text{e}}$, and $f_{\text{a}}$ for closeness, eccentricity, and all-subgraphs, respectively, are symmetric. Indeed, for all these functions a subtree \fix{always has less} potential than the whole tree. For this reason, $f_{\text{x}}(u, T_{u,v}) < f_{\text{x}}(v, T)$ for $\fix{\text{x}} \in \{\text{c}, \text{e}, \text{a}\}$ since $T_{u,v}$ is a subgraph of $T$. By Theorem~\ref{theo:rooting}, this proves that closeness, eccentricity, and all-subgraphs root trees. 
	
	In turn, the potential functions of degree centrality (as well as betweenness and decay centralities) are not symmetric and, therefore, do not root trees. For example, for degree centrality we can take $T = $ 
	\raisebox{-0.8mm}{\mbox{
\begin{tikzpicture}[node distance=2mm,text width=5mm]
\node (u) at (0,0) [stextnode] {u};
\node (u2) at (-0.5,0.13)  [snode] {};
\node (u3) at (-0.5,-0.13) [snode] {};

\draw [-, line width=0.2pt] (u) to (u2);
\draw [-, line width=0.2pt] (u) to (u3);

\node (u') at (0.6,0) [stextnode] {v};
%
\draw [-, line width=0.1pt] (u) to (u');

\end{tikzpicture}}}  
, and check that $f_{\text{d}}(u, T_{u,v}) > f_{\text{d}}(v, T)$. Therefore, $f_{\text{d}}$ is not symmetric. 
\end{example}


We showed that all standard centrality measures that \fix{root} trees can be defined through potential functions. The natural question is: is it true for all measures that root trees? In the following result, we show that this is not the case and there exists a measure \fix{that roots a tree}, but does not have a potential function.

\begin{proposition}\label{Prop:RootingNotPot}
	There exists a centrality measure that roots trees but does not admit a potential function.
\end{proposition}

Potential functions and Theorem~\ref{theo:rooting} explain why some measures root trees and others do not. In the following sections, we use this framework to further understand the tree rooting centrality measures in terms of algorithms, consistency, and the design of new measures.

\section{An algorithm to find the root}\label{Sec:alg}
%
%

Even though not every tree rooting measure has a potential function, having one gives us some essential properties that we can exploit. In particular, having a symmetric potential function implies finding the root of any tree in $\mathcal{O}(n \log(n))$-time \fix{under some assumptions on the efficiency to compute the potential function.} Notice that the naive approach of computing the centrality for each vertex separately and then choosing the one with the highest centrality runs in quadratic time (i.e., by assuming that computing the centrality of a single vertex takes linear time). Instead, the algorithm presented here runs in $\mathcal{O}(n \log(n))$ for every centrality measure that admits a symmetric potential function. 

The main intuition behind this algorithm is based on the following property satisfied by centrality measures with a symmetric potential function.

\begin{proposition}\label{prop:prev-algo}
Let $C$ be a centrality measure that has a symmetric potential function $f$. 
Let $T$ be any tree, $w_1,w_n \in V(T)$, and $w_1w_2...w_n$ be the unique path connecting $w_1$ to $w_n$ in~$T$.  Whenever $f(w_1, T_{w_1,w_2}) \leq f(w_n,T_{w_n,w_{n-1}})$ then $C(w_1, T) \leq C(w_2,T)$.
\end{proposition}

In other words, Proposition~\ref{prop:prev-algo} says that if the potential of the subtree hanging from $w_1$ is less than the potential of the subtree hanging from $w_n$, then a root should be closer to the adjacent vertex of $w_1$ that is towards the direction of $w_n$. This result gives us a way to traverse a tree, starting from the leaves and going up until we find the root. 
More specifically, starting from the leaves, by Proposition~\ref{prop:prev-algo} we can compare the potential of two opposite complete subtrees. Then, vertices with higher potential in their subtree indicate the direction of higher centrality. When we finally reach two connected vertices, the vertex with higher (or equal) potential is a root.

\newcommand{\qinsert}{\texttt{insert}}
\newcommand{\qpull}{\texttt{pull}}
\newcommand{\qq}{Q}

\begin{algorithm}[t]
	\setstretch{1.1}
	\DontPrintSemicolon
	\AlgoDontDisplayBlockMarkers
	\SetAlgoNoEnd
	\SetAlgoNoLine
	
	\SetKwFor{While}{while}{}{end while}
	\KwIn{A non-trivial tree $T$ and a symmetric potential function $f$.}
	\KwOut{The most central vertex of $T$ according to $f$.}
	
	\SetKwFunction{Root}{Find-a-root}
	\SetKwProg{Fn}{Function}{:}{}
	\Fn{\Root{$T, f$}}{
		
		$\qq \gets \texttt{Empty-queue}$\;
		$H \gets \{(v, N_T(v)) \mid v \in V(T)\}$\;
		
		\ForEach{$v$ leaf of $T$}{
			$\qq.\qinsert(v, f(v, G_v))$\;
		}
		\While{$Q.\texttt{size}() > 1$}{
			$v \gets \qq.\qpull()$\;
			
			$u \gets H[v]$ \;
			$H[u] \gets H[u] \setminus \{v\}$\;
			
			\If{$|H[u]| = 1$}
			{
				$w \gets H[u]$\;
				$\qq.\qinsert(u, f(u, T_{u,w}))$\;
			}	
		}
		\KwRet $\qq.\qpull()$\;
	}

	\caption{Find a root given a tree.}
	\label{Alg}
\end{algorithm}

Algorithm~\ref{Alg} implements the above intuition based on Proposition~\ref{prop:prev-algo}. It receives as input a tree $T$ and a symmetric potential function $f$, and outputs a root of $T$ with respect to $f$. For implementing Algorithm~\ref{Alg} we need two data structures, denoted by $H$ and~$Q$. The first data structure $H$ is a \emph{key-value map} (i.e., a Hash-table), where a key can be any vertex $v \in V(T)$ and its value is a subset of $N_T(v)$. We denote the value of $v$ (i.e., a key) in $H$ by $H[v]$. By some abuse of notation, when $H[v]$ is a single vertex, we write $u \gets H[v]$ to retrieve and store this vertex in $u$. The second data structure $Q$ is a \emph{priority-Queue}. For $v \in V(T)$ and $p \in \bbR$ we write $Q.\qinsert(v, p)$ to insert $v$ in $Q$ with priority $p$. We also write $v \gets Q.\qpull()$ to remove the vertex with the lowest priority from $Q$, and store it in $v$. For both structures, these operations can be implemented in $\mathcal{O}(\log(n))$-time where $n$ is the number of inserted objects~\cite{cormen2009introduction}. 

Algorithm~\ref{Alg} starts by initializing $Q$ as empty and $H$ with all key-value pairs $(v, N_T(v))$ (lines 2-3). Then, it runs over all leaves $v$ of $T$ and inserts it into $Q$ with priority $f(v, G_v)$ where $G_v$ is the tree with an isolated vertex $v$ (lines 4-5). As we already mentioned, the intuition is to start from all leaves $v$ and use its potential (as a single vertex) for comparing it with other vertices. Then we loop while the number of elements in $Q$ is greater than~$1$. Recall that any non-trivial tree has at least two leaves, and therefore the algorithm reaches line 6 with $Q.\texttt{size}() \geq 2$ for the first time. Instead, if $T$ is trivial, we return the single vertex directly in line 13.

We remove the vertex with the lowest priority from $Q$ in each iteration and store it in $v$ (line 7). This step discards $v$ as a possible root (by Proposition~\ref{prop:prev-algo}) and \fix{moves towards} its ``parent'' represented by $u \gets H[v]$ (line 8). Given that we discarded $v$, we remove $v$ as a child of $u$, where $H[u]$ contains the current children of $u$ (line 9). 
Indeed, when $|H[u]|=1$ (line 10) this means that we have reached $u$, its parent is $w \gets H[u]$ (line 11) and its complete subtree $T_{u,w}$ hanging from $u$ must be evaluated with $f$, and inserted in $Q$ (line~12). An important invariant during the while-loop is that any vertex $v$ in $Q$ satisfies $|H[v]| = 1$ (except at the end of the last iteration). Conceptually, if $H[v] = \{w\}$, this invariant means that $w$ is the parent of $v$ and we are using the potential of the subtree $(v, T_{v,w})$ for comparing $v$ with other vertices. Then when $v$ is the vertex with the lowest priority on $Q$, it means that other vertices beat it, and a root must be towards its parent. 

Finally, when there is only one vertex left in $Q$, it beats all other vertices, and it should be one of the roots. It is necessary to mention that if $T$ has two roots, we could also output the second root by slightly modifying the algorithm. 

Regarding time complexity, the reader can check that the for- and while-loops take linear time on $|T|$. Each operation over $H$ and $Q$ take at most $\log(|T|)$ steps, and overall it sum up to $\mathcal{O}(|T|\log(|T|))$ if computing $f$ takes constant time. 

Of course, the previous assumption is not always true, given that $f$ can be any symmetric potential function. To solve this, we say that $f$ is \emph{locally-computable} if, for every $T$ and $u \in V(T)$, $f(u,T)$ can be computed in $\mathcal{O}(k)$-time from the values of its $k$-neighbors, namely, from $f(u_1, T_{u_1,v}), \ldots, f(u_k, T_{u_k,v})$ where $N_T(u) = \{u_1, \ldots, u_k\}$. Note that by book-keeping the values $f(u_1, T_{u_1,v}), \ldots, f(u_k, T_{u_k,v})$ of the neighbors of $u$, we can compute $f(u, T_{u,w})$ (line 12) in $\mathcal{O}(|N_T(u)|)$-time. If we sum this extra time \fix{over all vertices}, it only adds $\mathcal{O}(|T|)$-steps to the total running time of the algorithm. 

\begin{proposition}\label{prop:AlgCorrect}
	Given a tree $T$ and a symmetric potential function $f$, Algorithm~\ref{Alg} returns a root of $T$ with respect to~$f$. Moreover, if $f$ is locally-computable, the algorithms runs in  $\mathcal{O}(|T|\cdot\log(|T|))$-time. 
\end{proposition}

The reader can easily check that the potential functions of closeness, eccentricity, and all-subgraphs are locally-computable (see also Section~\ref{Sec:alphabetagamma}), and then Algorithm~\ref{Alg} can be used for any of these measures to find the root of any tree in $\mathcal{O}(|T|\cdot\log(|T|))$.

\section{Consistent rooting}\label{Sec:regularity}
The tree rooting property fixes the ``shape'' of a centrality measure in every possible tree. 
However, it does not impose any relation between roots in different trees. 
\fix{As a result, even a small change (e.g., adding a leaf) may move the root arbitrarily since there might be no relation between the roots in a tree and the altered tree.}
To give an example, a centrality measure may be defined in one way for trees with odd number of vertices, but in a completely different way for trees with even number of vertices.

To this end, we propose a notion of \emph{consistency}.
Consistency states that if we add a leaf to the tree, then the root may move only in its direction.

\begin{definition}
We say that a centrality measure $C$ \emph{consistently roots trees} if it roots trees and for every tree $T$ and vertices $u,v \in V(T)$, $w \not \in V(T)$ such that $u \in \textsc{Max}_C(T)$ it holds $\textsc{Max}_C(T + \{v,w\}) \subseteq \pi_{u,w} \cup \textsc{Max}_C(T)$, where $\pi_{u,w}$ is the path between $u$ and $w$ in $T + \{v,w\}$.
\end{definition}

We can verify that closeness, eccentricity, and all-subgraphs centralities all consistently root~trees.

Consistency is a property of measures that root trees. 
However, it can be also interpreted using a natural property for arbitrary centrality measures that we call \emph{monotonicity}.
Monotonicity states that if vertex $v$ has a higher (or equal) centrality than its neighbour $u$ in a tree, then this fact will not change if we add a leaf on the side of vertex $v$.

\begin{definition}
We say that a centrality measure $C$ is \emph{monotonic} if for every tree $T$, vertices $v,u,w \in T$ such that $\{v,u\} \in E(T)$, $w \in T_{u,v}$ and vertex $w' \not \in V(T)$ if $C(v,T) < C(u,T)$, then $C(v,T + \{w,w'\}) < C(u,T+\{w,w'\})$.
\end{definition}


Monotonicity is in fact a general property satisfied by many centrality measures, including all geometric centralities such as closeness and decay.
The following result ties both concepts: monotonicity and consistency of rooting.

\begin{proposition}\label{Pr:ConsistencyChar}
Let $C$ be a centrality measure that roots trees. 
Then $C$ consistently \fix{roots} trees if, and only if, it is monotonic.
\end{proposition}

Let us turn our attention to the relation between tree rooting and potential functions. 
If a centrality that roots trees admits a potential function, then it must be consistent to some extent. 
However, as it turns out, it might not consistently root trees, as we show in the following counterexample.

\begin{example}\label{ex:regularity}
Consider the following ad-hoc centrality measure: 
\[ C(v,T) = \textsc{Eccentricity}(v,T) - (1/|T|^2) \cdot \textsc{Closeness}(v,T). \]
Intuitively, if two vertices in a tree have different eccentricity, then their eccentricity differs by more than $1/|T|^2$.
Also, $\textsc{Closeness}(v,T) \in (0,1]$. 
Hence, we have $C(u,T) < C(v,T)$ if, and only if, $u$ has a smaller eccentricity than $v$ or equal eccentricity, but higher closeness.

It is easy to verify that the following potential function corresponds to $C$: $f(v,T) = h(v,T)-1/|T|$, where $h(v,T)$ is the distance from $v$ to the farthest vertex in $T$.

To show that consistency is violated consider trees \raisebox{-0.8mm}{\mbox{
\begin{tikzpicture}[node distance=2mm,text width=5mm]
\node (u) at (0,0) [stnode] {};
\node (u2) at (-0.3,-0.13)  [snode] {};

\draw [-, line width=0.1pt] (u) to (u2);

\node (u') at (0.3,0) [snode] {};
\node (u2') at (0.6,0.13)  [snode] {};
\node (u3') at (0.6,-0.13) [snode] {};

\draw [-, line width=0.1pt] (u') to (u2');
\draw [-, line width=0.1pt] (u') to (u3');
\draw [-, line width=0.1pt] (u) to (u');

\end{tikzpicture}}}  
 and \raisebox{-0.8mm}{\mbox{
\begin{tikzpicture}[node distance=2mm,text width=5mm]
\node (u) at (0,0) [stnode] {};
\node (u2) at (-0.3,-0.13)  [snode] {};
\node (u3) at (-0.3,0.13)  [snode] {};

\draw [-, line width=0.1pt] (u) to (u2);
\draw [-, line width=0.1pt] (u) to (u3);

\node (u') at (0.3,0) [stnode] {};
\node (u2') at (0.6,0.13)  [snode] {};
\node (u3') at (0.6,-0.13) [snode] {};

\draw [-, line width=0.1pt] (u') to (u2');
\draw [-, line width=0.1pt] (u') to (u3');
\draw [-, line width=0.1pt] (u) to (u');

\end{tikzpicture}}}  
 (vertices with the highest centrality are marked with white color). 
Consistency states that in the second tree the root should stay on the left-hand side which is not the case here.
\end{example}

To characterize which of the centrality measures with potential function root trees consistently, we look at the restriction that monotonicity imposes on the potential function.

\begin{proposition}\label{prop:monotonicity_potential}
Let $C$ be a centrality measure that admits a potential function $f$.
Then $C$ is monotonic iff for every tree $T$, subtree $T'$ of $T$, and $v \in V(T')$ it holds~$f(v,T) \ge f(v,T')$.
\end{proposition}

In particular, the potential function from Example~\ref{ex:regularity} violates this condition, as adding vertices to a tree without increasing its height decreases the value of the potential function.

Now we can summarize rooting results and consistency regarding potential functions as one of our principal theorems.

\begin{theorem}\label{prop:MonotonicSymmetric}
Let $C$ be a centrality measure that admits a potential function $f$.
Then $C$ is monotonic and symmetric if, and only if, for every tree $T$, proper subtree $T'$ of $T$ and vertices $v \in V(T)$, $u \in V(T')$ it holds $f(v,T) \ge f(u,T')$ and $f(v,T) > f(u,T')$ if $u \neq v$.
\end{theorem}

\section{Families of potential functions}\label{Sec:alphabetagamma}

In this section, we apply the previous results by showing how to design potential functions that consistently root trees. Specifically, using the following results, we can derive an infinite family of potential functions. This family shows infinite ways to root trees with good characteristics, namely, that are consistent and computable in $\mathcal{O}(n \log(n))$ time. In the following, we recall some standard definitions of monoids, to then define potential functions through them. 


A \emph{monoid} (over $\bbR$) is a triple $(\mdom, \mop, \mone)$ where $\mdom \subseteq \bbR$, $\mop$ is a binary operation over $\mdom$, $\mop$ is associative, and $\mone \in \mdom$ is the identity of $\mop$ (i.e., $r \mop \mone = \mone \mop r = r$). We further assume that monoids are commutative, namely, $\mop$ is commutative. Examples of (commutative) monoids are $(\bbR_{\geq 0}, +, 0)$ and $(\bbR_{\geq 1}, \times, 1)$, where we use $\bbR_{\geq c}$ for all reals greater or equal than~$c$. For the sake of presentation, in the following we will usually refer the monoid 

\begin{definition}
Given a monoid $(\mdom, \mop, \mone)$ and $\ell: \mdom \rightarrow \mdom$, we define the potential function $\cpf$ recursively as follows:
\begin{enumerate}
	\item For a vertex $v$, we define $\cpf(v, G_v) = \mone$, where $G_v$ is the tree with an isolated vertex $v$.
	
	\item For a tree $T$ and a leaf $v \in V(T)$ hanging from its parent $u \in V(T)$, we define $\cpf(v, T) = \ell(\cpf(u, T_{u,v}))$. In other words, we apply $\ell$ to the potential function of the subtree rooted at $u$. We call $\ell$ the \emph{leaf-function} of $\cpf$. 
	
	\item Given two trees $T_1$ and $T_2$ with $V(T_1) \cap V(T_2) = \{v\}$, we define $\cpf(v, T_1 \cup T_2) = \cpf(v, T_1) \mop \cpf(v, T_2)$. 
\end{enumerate}
A potential function $f$ is \emph{constructive} if there exists a monoid $\mop$ and a leaf-function $\ell$ such that $f = \cpf$.
\end{definition}
Notice that $\cpf(v, T)$ is uniquely determined by the three cases above. Specifically, suppose that $u_1, \ldots, u_k \in N_T(v)$ are the neighbors of $v$ on $T$. Then we can decompose $T$ by considering all subtrees $T_{u_i,v} + \{u_i,v\}$ and compute $\cpf(v, T)$ recursively:
$$
\cpf(v, T) = \ell\big(\cpf(u_1, T_{u_1,v})\big) \mop \ldots \mop \ell\big(\cpf(u_k, T_{u_k,v})\big)
$$
until we reach a single vertex. Furthermore, $\cpf$ is closed under isomorphism over $\cN$ given that $\mop$ is associative and commutative. Thus, we conclude that $\cpf$ is well-defined and could work as a potential function. In addition, $\cpf$ is locally-computable since $\cpf(v, T)$ can be computed from its $k$-neighbors.

\begin{example}\label{ex:constructive}
All potential functions presented in Example~\ref{ex:pot-func} are constructive by considering the following monoids and leaf-functions:
$$
\begin{array}{lll}
	(\bbR_{\geq 0}, +, 0) & \ell(x) = 1  & (degree)\\
	(\bbR_{\geq 1}, a+b-1, 1) & \ell(x) = x+1 & (closeness)\\
	(\bbR_{\geq 0}, \max, 0) \ \ & \ell(x) = x+1 \ \  & (eccentricity) \\
	(\bbR_{\geq 1}, \times, 1)  & \ell(x) = x+1  & (all\text{-}subgraphs)
\end{array}
$$
It is easy to check that each monoid and leaf-function defines the corresponding potential function of the above measures. 
\end{example}

One advantage of the previous definition is that it shows a way for constructing potential functions. Moreover, we can study which properties are necessary over $\mop$ and $\ell$ to guarantee that $\cpf$ consistently root trees. Towards this goal, we recall some standard definitions for monoids and functions. A function $f$ is called \emph{monotonic} if $x \leq y$, implies $f(x) \leq f(y)$ for every $x, y$. A monoid $(\mdom, \mop, \mone)$ is called \emph{ordered} if $x \leq y$, implies $x \mop z \leq y \mop z$ for every $x, y, z \in \mdom$. Further, it is called \emph{positively ordered} if in addition $\mone \leq x$ for every $x \in \mdom$. 

\begin{lemma}\label{lemma:const}
	Let $(\mdom, \mop, \mone)$  be a monoid and $\ell: \mdom \rightarrow \mdom$ a leaf-function. The potential function $\cpf$ consistently roots trees whenever (1) $x < \ell(x)$ for every $x$, (2) $\ell$ is monotonic, and (3)~$(\mdom, \mop, \mone)$ is positively ordered. 
\end{lemma}

For example, the monoids and leaf-functions of closeness, eccentricity, and all-subgraphs (see Example~\ref{ex:constructive}), \fix{satisfy} properties (1) to (3) and, as we know, they consistently root trees. On the other hand, degree's leaf-function does not satisfy (1), and therefore, it does not root trees.  

Lemma~\ref{lemma:const} shows sufficient conditions over $\mop$ and $\ell$ to consistently root trees. To get a necessary condition, we need to add some technical restrictions, and to slightly weaken conditions (2) and (3). Towards this goal, let $\cpfran$ be the range of $\cpf$. Define $\bar{\mop}$ and $\bar{\ell}$ to be the monoid $(\mdom, \mop, \mone)$ and function $\ell$ restricted to $\cpfran$. For two values $x$ and $y$, we say that $x$ is a \emph{subtree-value} of $y$ if there exist $T$ and $T'$ such that $T$ is a subtree of $T'$, $\cpf(u, T) = x$, and $\cpf(u, T') = y$ for some $u\in V(T)$. Then we say that $\bar{\ell}$ is \emph{monotonic over subtrees} if $x \leq y$ and $x$ is a subtree-value of $y$ implies that $\bar{\ell}(x) \leq \bar{\ell}(y)$. Similarly, we say that $\bar{\mop}$ is \emph{positively ordered over subtrees}, if  $x \leq y$ and $x$ is a subtree-value of $y$, then $x \mop z \leq y \mop z$ for every $z \in \cpfran$, and $\mone \leq x$ for every $x \in \cpfran$.

\begin{theorem}\label{theo:const}
	Let $(\mdom, \mop, \mone)$  be a monoid and $\ell: \mdom \rightarrow \mdom$ a leaf-function. The potential function $\cpf$ consistently roots trees if, and only if, (1) $x < \bar{\ell}(x)$ for every $x \in \cpfran$, (2) $\bar{\ell}$ is monotonic over subtrees, and (3)~$\bar{\mop}$ is positively ordered over subtrees.
\end{theorem}

Theorem~\ref{theo:const} and, specifically, Lemma~\ref{lemma:const} give the ingredients to design potential functions that consistently root trees and, further, we have an algorithm to find the root in $\mathcal{O}(n \log(n))$. 
For instance, take a triple $(a,b,c) \in \bbR^3$. Then define the monoid $(\bbR_{\geq c}, \mop_c, c)$ and leaf-function $\ell_{a,b}$ such that:
$
x \mop_c y :=  \frac{x \cdot y}{c}$ \, and \, $\ell_{a,b}(x) := a\cdot x + b
$.
For example, if we consider $a=b=c=1$, we get the monoid and leaf-function for the potential function of all-subgraph centrality (see Example~\ref{ex:constructive}). 
Interestingly, one can verify that, if $a \geq 1$, $b > 0$ and $c>0$, then $\mop_c$ is a monoid. Moreover, $\mop_c$ and $\ell_{a,b}$ satisfy properties (1) to (3) of Lemma~\ref{lemma:const} and we get the following result.

\begin{proposition}\label{Lemma:alphabetagamma}
	For every $(a, b, c) \in \bbR^3$ with $a \geq 1$, $b >0$, and $c > 0$, the potential function $f_{\mop_{c},\ell_{a,b}}$ consistently root trees.
\end{proposition}

Finally, we want to know if we can get different roots for different values $(a,b,c)$. In other words, is it the case that for every different triples $(a,b,c)$ and $(a',b',c')$ there exists a tree $T$ such that the root of $T$ according to $f_{\mop_c,\ell_{a,b}}$ is different to one chosen by $f_{\mop_{c'},\ell_{a', b'}}$? The next result shows that $\{f_{\mop_{c},\ell_{a,b}} \mid c \geq 1, b>0,c >0\}$ is indeed an infinity family of different potential functions for tree rooting. 

\begin{proposition}\label{prop:InfiniteRooting}
	There exists an infinite set $S \subseteq \bbR^3$ such that for every $(a,b,c), (a',b',c')\in S$, there exists a tree $T$ where the roots of $T$ according to $f_{\mop_c,\ell_{a,b}}$ are not the same as roots of $T$ according to~$f_{\mop_{c'},\ell_{b',c'}}$.
\end{proposition}

\section{Discussion}\label{Sec:concl}
\fix{We end the paper by discussing some extensions and applications.}

\paragraph{Extensions to other classes of graphs} \fix{A natural criticism of focusing on trees is whether one can generalize the know-how acquired on trees to other classes of graphs. We agree that further research is needed to extend potential functions to new graph families. Nevertheless, we see some exciting directions in which our work can be extended. In particular, the idea of potential functions extends to arbitrary graphs by considering the endpoints of a bridge instead of adjacent vertices of a tree.
More in detail, let $G$ be a connected graph and $\{u,v\}$ be an edge such that $G_u,G_v$ are two connected components of $G - \{u,v\}$ that contain $u,v$, respectively.
We say that $f: \cN \to\mathbb{R}$ is a \emph{graph} potential function for $C$ if for every such graph $G$ it holds $C(u,T) \leq C(v,T)$ if, and only if, $f(u,G_u) \leq f(v,G_v)$.
This property generalizes (tree) potential functions, as every two adjacent nodes in a tree form a bridge.
As it turns out, all centrality measures listed on Figure~\ref{fig:symmetry} that have (tree) potential functions (degree, closeness, betweenness, eccentricity, all-subgraphs, decay) have identical graph potential functions.
For example, $\textsc{Eccentricity}(v,G) \le \textsc{Eccentricity}(u,G)$ if and only if $f_e(v,G_v) \le f_e(u,G_u)$ for $f_e$ defined in Example~\ref{ex:pot-func}.
Interestingly, some centrality measures have identical potential functions on trees but different ones on general graphs (for example, closeness and random-walk closeness centralities).}

\paragraph{Applications} \fix{In this work, we focused on the foundational aspects of understanding centrality measures over trees, and we left for future work the application in the context of data management. Given the axiomatic approach of our work and given that tree structures are ubiquitous in data management, we believe that potential functions and their implications on rooting trees could find several exciting applications. In the following, we present some possible applications of this work in data management scenarios.}

\fix{In conjunctive query answering, the class of acyclic queries is of particular interest, given that each query has a join tree that permits efficient evaluation in linear time on data complexity~\cite{abiteboul1995foundations}. For this, the so-called Yannakakis algorithm~\cite{yannakakis1981algorithms} performs a bottom-up traversal of the join tree for filtering the tuples that will not be part of the output. In particular, the different ways one can root the join tree gives rise to several individual computational schedules to obtain the same results~\cite{abiteboul1995foundations}. Here, rooting the join tree could be exploited to improve existing join evaluation algorithms by using a particular potential function that uses the query structure and database relations. We leave for future work on how one can use this principle for query evaluation in the presence of join trees.}

\fix{Another possible application is in the context of tree-structured data, like XML or JSON documents. Although this data is usually rooted, assessing the most crucial node using a centrality measure can lead to a better understanding of the document's structure. For instance, given a tree-structured document, one could measure the difference between the root provided by potential function and the original root and see how this difference affects query evaluation, document representation, or other metrics.}

\fix{Finally, in a broader sense, one could see centrality measures over graphs as an instance of a general database problem: find the most central data object in the data model given its underlying structure. The data model could be a relational database, an object-oriented database, an RDF database, or even a tree-structure database. For all these cases, the principle should be the same: the more relevant the data object is for its data model, the more central it should be. The present work could be seen as the first step toward this direction, namely, understanding data centrality in the case of trees. We leave for future work on how to extend this line of research to other classes of graphs or data models.}

\bibliography{references}

\newpage
\appendix

\section{Other examples of centrality measures}

\paragraph{Betweenness centrality} For every graph $G$ and vertices $v,u$ in $V(G)$ we define the set of geodesic paths between $u$ and $v$ as $D_{vu} = \{\pi \subseteq G \mid \pi \text{ is a shortest path from }v \text{ to } u\}$. Analogously, for three vertices $v,u,w$, we define the set of geodesic paths from $v$ to $u$ containing $w$ as $D_{vu}(w)= \{\pi \in D_{vu}\mid w\in V(\pi)\}$. Thus, \emph{betweenness centrality}~\cite{freeman1977set} of $v$ in $G$ is defined as $$\textsc{Betweenness}(v,G) = \sum_{(u,w)\in (V(G)-\{v\})^{2}}\displaystyle\frac{|D_{uw}(v)|}{|D_{uw}|}.$$
In some sense, betweenness centrality is based on the notion of distance. However, we do not assign importance to the distance itself but to the structures where vertex $v$ belong. This measure is closely related with connectivity or flux between vertices. 

\paragraph{Decay centrality} Distance is relevant in several centrality measures. In particular, we can define the $i$th neighbourhood of a node $v$ in the graph $G$ as $N^{i}_G(v) = \{u\in V(G)\mid d_G(u,v) = i\}$. In some applications, we would like to give more importance to a node with more closer nodes to him. This is essentially the idea of  $\alpha-$\text{\it Decay centrality}~\cite{jackson2010social}. We define it as
$$
\textsc{Decay}^{\alpha}(v,G) = \sum_{i=1}^{|V(G)|} \alpha^{i}|N^{i}(v,G)| \text{ for }\alpha\in(0,1).
$$
Clearly, for every $\alpha$ in $(0,1)$ we might get a different centrality measure. Therefore, we say that $\alpha-$Decay is a family of centrality measures.

\paragraph{Eigenvector centrality}  For every graph $G$, given an order of the nodes $V(G)= \{u_1,...,u_{|V(G)|}\}$ we define the adjacency matrix of $G$ as $A_{ij}=1$ if $\{u_i,u_j\}\in E(G)$. 
Then, $\lambda_{\max}$ is the greatest eigenvalue of $A$.
We define the unique $L^2$ normalized eigenvector associated with $\lambda_{\max}$ as $\mathbf{v}_{\max}$. Thus, the \emph{eigenvector centrality}~\cite{bonacich1972factoring} of $u_i$ in $G$ is the $i$-th entrance of $\mathbf{v}_{\max}$: $$\textsc{EV}(u_i,G)= \mathbf{v}_{\max}^{i}.$$

\paragraph{PageRank centrality}  \emph{PageRank centrality} is one of the main algorithms used in Google's web searching engine. For a graph $G$, define $\textbf{P}$ as the column-stochastic matrix such that:
$$
\textbf{P}_{ij} = \displaystyle\frac{A_{ij}}{|N(i,G)|}.
$$
In other words, $\textbf{P}$ contains the probabilities of jumping from node $i$ to node $j$ during a random walk over $G$. Let $\textbf{v}$ be an stochastic vector $(e^T\textbf{v} = 1)$, and let $0 < \alpha < 1$ be a teleportation parameter. Then, $\textsc{PageRank}$ is defined as the vector who solves the equation $(I-\alpha \textbf{P})\textsc{PageRank} = (1-\alpha)\textbf{v}$. Finally, we define \emph{PageRank centrality}~\cite{page1999pagerank} as $$\textsc{PR}_{\alpha,\textbf{v}}(v_i,G) = \textsc{PageRank}_{i}.$$

\section{Proofs of Section~\ref{Sec:rooting}}
\subsection*{Proof of Proposition~\ref{prop:amoproot}}
\begin{proof}
$(\Rightarrow)$ Suppose $C$ satisfies the rooting property. By contradiction, suppose there exists a tree $T$ and node $v$ in $T$ with two neighbours in $u_1\not = u_2$ in $N_{T}(v)$ where $C(v, T)\leq C(u_1, T)$ and $C(v, T) \leq C(u_2)$. Set one root of $T$ according to $C$ as $r$ in $V(T)$. Now, without lose of generality, assume that $\pi_{v r} = v u_1...r$. In other words, from $v$ to $r$ we need to pass through $u_1$. This must be the case since $r\not=v$. Otherwise, $v,u_1$ and $u_2$ are roots which is not possible. Now, the unique path from $u_2$ to $r$ must go through $v$ and $u_1$. However, the centrality values are not strictly increasing as it should be for a rooting measure.

\smallskip
\noindent $(\Leftarrow)$ In first place, we will prove that for every tree $T$, $v$ in $V(T)-\maxcen(T)$ with a closest root $r\in\maxcen(T)$, then if $\pi_{rv} = u_1u_2...u_n$ we have that $C(u_1,T)>C(u_2,T)>...>C(u_n,T)$. Since $r$ is the closest root to $v$ in $T$, then necessarily $C(u_1, T) = C(r, T) > C(u_{2})$. Now, suppose as induction hypothesis that $C_(u_i, T) > C(u_{i+1}, T)$. Then, by at most one parent we have that $C(u_{i+2}, T) < C(u_{i+1}, T)$. Otherwise $u_{i+1}$ would have two neighbours with higher or equal centrality. 

In second place, in order to prove that the set $\maxcen(T)$ forms a clique in $T$. Consider two roots $u,v$ in $T$. By contradiction suppose $\{u,v\}\not\in E(T)$. Then, set the path connecting both nodes in $T$ as $\pi_{vu} =  w_1w_2...w_n$ where $n \geq 3$ because of our assumption. Now, we have two options. First it could be the case that for every $w_i\not\in\{u,v\}$ then $w_i$ is not a root of $T$. Then, we know the centrality values must decrease and then increase, which violates what we prove before. On the other hand, if there exists $w_i\not\in\{u,v\}$ such that $w_i$ is a root of $T$ the same scenario will occur, violating the at most one parent property. This prove that from every two roots in $T$, they must be connected by an edge which concludes the proof.
\end{proof}
\subsection*{Proof of Proposition~\ref{pr:RootSym}}

\begin{proof}
	Suppose by contradiction there exist a tree $T$, a node $v$ in $V(T)$ and two different nodes $u_1$ and $u_2$ in $V(T)$ such that $(u_1, T_{u_{1} v})\cong (u_2, T_{u_{2} v})$ but $C(v, T) \leq C(u_1, T)$. This means that there exists a node $z$ in $V(T_{u_{1}v})$ such that $z\in\maxcen(T)$ because the centrality is increasing in the direction of $u_1$. However, since $T_{u_{1}v}\equiv T_{u_{2}v}$, there must exists a node $z'$ in $V(T_{u_{2}v})$ such that $C(z, T) = C(z', T)$ for the close under isomorphism property of $C$. In consequence, we have that $z'$ is in $\maxcen(T)$ as well. This means there are two roots of $T$ which are not connected, but at the same time $C$ is a tree rooting measure.
\end{proof}

\section{Proofs of Section~\ref{Sec:potential}}
\subsection*{A potential function for \textsc{Betweenness} (Example~\ref{ex:pot-func})}

\begin{proof}
	Take any two nodes $v_1$ and $v_2$ connected in a tree $T$. By definition of \textsc{Betweenness} we have that $$
	\textsc{Betweenness}(v_1, T) = \displaystyle\sum_{(u,w)\in (V(T)-\{v_1\})^{2}}\displaystyle\frac{D_{uw}(v_1)}{|D_{uw}|} =$$ $$ \displaystyle\sum_{(u,w)\in (V(T_{v_1v_2})-\{v_1\})^{2}}\displaystyle\frac{D_{uw}(v_1)}{|D_{uw}|} + \displaystyle\sum_{(u,w)\in (V(T_{v_1v_2})-\{v_1\})\times V(T_{v_2v_1})}2 \cdot \displaystyle\frac{D_{uw}(v_1)}{|D_{uw}|}.
	$$
	In other words, the shortest paths containing $v_1$ can be divided in two groups. First, the ones connecting nodes inside $T_{v_1v_2}$ and second, the ones connecting nodes from $T_{v_1v_2}$ to $T_{v_2v_1}$. In the first group, we are exactly computing the betweenness value for $v_1$ in tree $T_{v_1v_2}$. Meanwhile, for the second group, we have one shortest path for every possible pair selection since we are force to pass through $v_1$. Therefore, we can express the previous value as $$\textsc{Betwenness}(v_1, T) = \textsc{Betwenness}(v_1, T_{v_1v_2}) + 2(|T_{v_1v_2}|-1)|T_{v_2v_1}| =$$ $$\textsc{Betwenness}(v_1, T_{v_1v_2}) + 2|T_{v_1v_2}||T_{v_2v_1}| - 2|T_{v_2v_1}|.$$
	Finally, we can use this expression to prove that $\textsc{Betweenness}(v_1, T) \leq \textsc{Betweenness}(v_2, T)$ if and only if $$\textsc{Betwenness}(v_1, T_{v_1v_2}) + 2|T_{v_1v_2}||T_{v_2v_1}| - 2|T_{v_2v_1}|\leq \textsc{Betwenness}(v_2, T_{v_2v_1}) + 2|T_{v_2v_1}||T_{v_1v_2}| - 2|T_{v_1v_2}|.$$
	Rearranging and simplifying terms we get that $\textsc{Betweenness}(v_1, T) \leq \textsc{Betweenness}(v_2, T)$ if and only if $$\textsc{Betwenness}(v_1, T_{v_1v_2}) + 2|T_{v_1v_2}|\leq \textsc{Betwenness}(v_2, T_{v_2v_1}) + 2|T_{v_2v_1}|.$$
	Which proves that $f_B(v, T) = \textsc{Betweenness}(v, T) + 2|T|$ is a potential function for $\textsc{Betwenness}$. Clearly, $f_B$ is not symmetric since a leaf will generally get less value than its parent before adding such leaf (as long as the parent was not a leaf before).
\end{proof}

\subsection*{A potential function for $\textsc{Decay}^{\alpha}$ (Example~\ref{ex:pot-func})}

\begin{proof}
	In the same way we did for other centrality measures, for a pair of connected nodes $v_1$ and $v_2$ inside $T$, we have the following relation: $$\textsc{Decay}^{\alpha}(v_1, T) = \sum_{i=1}^{|V(T_{v_1v_2})|}\alpha^{i}|N^{i}(v_1, T_{v_1v_2})| + \sum_{i=0}^{|V(T_{v_2v_1})|}\alpha^{i+1}|N^{i}(v_2, T_{v_2v_1})| =$$ $$ \textsc{Decay}^{\alpha}(v_1, T_{v_1v_2}) + \alpha(1+\textsc{Decay}^{\alpha}(v_2, T_{v_2v_1})).$$
	To derive this expression we just divide the neighbourhoods $N^{i}(v_1, T)$ in two groups. First, nodes coming from $T_{v_1v_2}$ and nodes from $T_{v_2v_1}$ in the other hand. Using this equation and its symmetry for $v_2$, we have that $\textsc{Decay}^{\alpha}(v_1, T) \leq \textsc{Decay}^{\alpha}(v_2, T)$ if and only if $$\textsc{Decay}^{\alpha}(v_1, T_{v_1v_2}) + \alpha(1+\textsc{Decay}^{\alpha}(v_2, T_{v_2v_1}))\leq \textsc{Decay}^{\alpha}(v_2, T_{v_2v_1}) + \alpha(1+\textsc{Decay}^{\alpha}(v_1, T_{v_1v_2})),$$
	by rearranging terms we have that this inequality holds if and only if $$(1-\alpha)\textsc{Decay}^{\alpha}(v_1, T_{v_1v_2}) \leq (1-\alpha)\textsc{Decay}^{\alpha}(v_2, T_{v_2v_1}).$$
	In other words, the function $f_{\alpha}(v, T) = (1-\alpha)\textsc{Decay}^{\alpha}(v, T)$ is a potential function for $\alpha-$\textsc{Decay} centrality. It is important to notice that $f_{\alpha}$ is a family of potential functions since it depends on how we choose $\alpha$. Finally, this potential function is clearly not symmetric for a similar reason to \textsc{Betweenness}.
	 
\end{proof}
\subsection*{Proof of Lemma~\ref{Lemma:SymmetricPot}}

\begin{proof}
$(\Rightarrow)$ Take a tree $T'$ and node $u'$ in $T'$ such that $(u, T_{u,v})\equiv (u', T')$. Know, by symmetry, we know that $C(v, T+T'+\{v,u'\}) > C(u', T+T'+\{v,u'\})$. Now, by potential functions definition that $f(v, T) > f(u', T') = f(u, T_{u,v})$ which we can conclude by close under isomorphism.

\smallskip
\noindent $(\Leftarrow)$ By counter positive suppose there exists a tree $T$ and nodes $v,u$ in $T$ such that $f(u, T_{u,v})\geq f(v, T)$. Then, we could connect to $v$ in $T$ a tree $T'$  such that for $u'$ in $T'$ we have that $(u, T_{u,v})\equiv (u', T')$. In this case, because of isomorphism, we know that $f(u', T') = f(u, T_{u,v}) \geq f(v, T)$. Now, by definition of potential functions, we have that $C(u', T+T'+\{u',v\}) \geq C(v, T+T'+\{u',v\})$ concluding that $C$ is not symmetric.
\end{proof}

\fix{
\subsection*{Proof of Theorem~\ref{theo:rooting}}
\begin{proof}
	We know by Proposition~\ref{pr:RootSym} that if $C$ roots trees, then $C$ is necessarily symmetric. Thus, we just need to prove that symmetry implies tree rooting for $C$. We will do this by proving that $C$ satisfies \fix{the} at-most-one-parent property. Fix an arbitrary tree $T$, vertex $v$ and let $u,w$ be two arbitrary \fix{neighbours} of $v$ in $T$.
	By removing edges $\{u,v\}$ and $\{v,w\}$ we get a decomposition of $T$ into trees, $T_u, T_v, T_w$ containing $u,v,w$, respectively, such that
	\[ T = (T_v \cup T_u \cup T_w)+ \{u,v\} + \{v,w\}. \]
	Without loss of generality assume $f(w,T_w) \leq f(u,T_u)$.
	From symmetry we have $f(u,T_u) < f(v, (T_u + \{u,v\}) \cup T_v)$.
	Based on our assumption $f(w,T_w) \leq f(u,T_u)$, we conclude  $f(w,T_w) < f(v, (T_u + \{u,v\}) \cup T_v)$.
	This, combined with the potential function definition, implies that $C(w,T) < C(v,T)$.
	Hence, we just proved that out of every two neighbours at least one has smaller centrality than $v$, which implies the thesis.
\end{proof}
}

\subsection*{Proof of Proposition~\ref{Prop:RootingNotPot}}

\begin{proof}
	In first place we will determine a fixed way to set the root of every tree defined as a function $R:\mathcal{G}\to \mathcal{V}$ such that $R(T)\subset V(T)$. For an  arbitrary tree $T$, we define $R(T)$ in cases as follows:
	
	\begin{enumerate}
		\item If $\arg\max_{v\in V(T)}\textsc{Eccentricity}(v,T) = \{w_1,w_2\}$ then $R(T) = \arg\max\{|T_{w_1,w_2}|,|T_{w_2,w_1}| \}.$ In other words, if there are two roots according to $\textsc{Eccentricity}$ we choose the one with bigger subtree. In case we have a tie, we just keep both nodes as roots.
		
		\item If $\arg\max_{v\in V(T)}\textsc{Eccentricity}(v,T) = \{w\}$, define the set $M(T,w) = \arg\max_{u\in N_{T}(w)}|T_{u,w}|$. If $|M(T,v)| = 1 $ we define $ R(T) = M(T, w)\cup \{w\}$. Otherwise, we just set $R(T) = \{w\}$. In simpler words, if there is just one root according to $\textsc{Eccetricity}$ we choose this root plus the child with bigger subtree just in case there is a unique dominant child.
	\end{enumerate}

	Finally, we define $$C^{\star}(v, T)=\begin{cases}
		2 & \text{ if }v\in R(T)\\
		\displaystyle\max_{u\in R(T)}\frac{1}{d_{T}(v,u)} & \text{otherwise}
	\end{cases}.$$

Clearly, $C^{\star}$ is a tree rooting measure since the most important node is based on eccentricity and centrality values from the root decrease. Thus, we will prove that $C^{\star}$ can not have a potential function. Consider the rooted trees $(v_1,T_1)=$\raisebox{-0.8mm}{\mbox{
		\begin{tikzpicture}[node distance=2mm,text width=5mm]
			\node (u) at (0,0) [stnode] {};
			\node (u2) at (-0.3,-0.13)  [snode] {};
			\node (u3) at (-0.3,0)  [snode] {};
			\node (u') at (-0.3,0.13) [snode] {};
			
			\draw [-, line width=0.1pt] (u) to (u2);
			\draw [-, line width=0.1pt] (u) to (u3);
			\draw [-, line width=0.1pt] (u) to (u');

\end{tikzpicture}}}  
,
$(v_2,T_2)=$\raisebox{-0.8mm}{\mbox{
		\begin{tikzpicture}[node distance=2mm,text width=5mm]
			\node (u) at (0,0) [stnode] {};
			\node (u2) at (-0.3,0)  [snode] {};
			
			\draw [-, line width=0.1pt] (u) to (u2);
			
			\node (u') at (0.3,0) [snode] {};
			\node (u2') at (0.6,0)  [snode] {};
			
			\draw [-, line width=0.1pt] (u') to (u2');
			\draw [-, line width=0.1pt] (u) to (u');
			
\end{tikzpicture}}}  
 and
$(v_3,T_3)=$\raisebox{-0.8mm}{\mbox{
		\begin{tikzpicture}[node distance=2mm,text width=5mm]
			\node (u) at (0,0.3) [snode] {};
			\node (u2) at (-0.3,0.3)  [stnode] {};
			
			\draw [-, line width=0.1pt] (u) to (u2);
			
			\node (u') at (0.3,0.3) [snode] {};
			\node (u2') at (0.6,0.3)  [snode] {};
			
			\draw [-, line width=0.1pt] (u') to (u2');
			\draw [-, line width=0.1pt] (u) to (u');
			
\end{tikzpicture}}}  
. The reader can corroborate that $C^{\star}(v_1, T_1+T_2 +\{v_1,v_2\})=C^{\star}(v_2, T_1+T_2 +\{v_1,v_2\})$, $C^{\star}(v_2, T_2+T_3+\{v_2,v_3\}) = C^{\star}(v_3, T_2+T_3+\{v_2,v_3\})$ and $C^{\star}(v_1, T_1+T_3+\{v_1,v_3\}) < C^{\star}(v_3, T_1+T_3+\{v_1,v_3\})$. Now, suppose by contradiction there exists a potential function $f$ for $C^{\star}$. Therefore, the previous values for $C^{\star}$ would mean that $f(v_1, T_1) = f(v_2, T_2) = f(v_3, T_3)$ but the fact that $C^{\star}(v_1, T_1+T_3+\{v_1,v_3\}) < C^{\star}(v_3, T_1+T_3+\{v_1,v_3\})$ means that $f(v_1, T_1) < f(v_3, T_3)$ which is a contradiction.
\end{proof}

\section{Proofs of Section~\ref{Sec:alg}}

\subsection*{Proof of Proposition~\ref{prop:prev-algo}}
\begin{proof}
	By potential function definition we know that 
	$$
		C(u, T) \leq C(w_2,T) \text{ if, and only if, } f(u, T_{u,w_2}) \leq f(w_2,T_{w_2,u}).
	$$
	In consequence, we just have to prove the second inequality in the previous equation. Now, by symmetry
	$$
		f(w_2,T_{w_2,u}) > f(w_3, T_{w_3,w_2}). 
	$$
	We can repeat this until we reach $w_{n} = v$ and then 
	$$
		f(w_2,T_{w_2,u}) > f(v, T_{v,w_{n-1}}).
	$$
	Finally, by hypothesis we know that  $f(v, T_{v,w_{n-1}}) > f(u, T_{u,w_2})$. Thus, this fact plus the last equation implies that
	$$
		f(w_2, T_{w_2,u}) > f(u, T_{u,w_2}).
	$$
\end{proof}

\subsection*{Proof of Proposition~\ref{prop:AlgCorrect}}

\begin{proof}
	
	First we will prove that for every input $T,f$ such that $f$ is a symmetric potential function, then the algorithm finish in finite time. We will prove that for every pair  $(u,f(u,T_{u,w}))\in Q$, then $|H(v)| = 1$. In the first iteration, for every leaf, it is clearly the case. Now, suppose this assumption is true at the beginning of one iteration of the while loop. Therefore, we have that $(u, f(u,T_{u,w}))$ enters the queue only if $|H[u]| = 1$ concluding this invariant for every iteration. On the other hand, if $v$ leaves the queue, then necessarily $H[v]$ is more important than $v$ in $T$. Since $v$ left the queue, there exists $w\in Q$ such that $f(w, T_{w,H(w)}) \geq f(v,T_{v,H[v]})$. Now, it is the case that $H[v]$ is in the path from $v$ to $w$. To prove this, by induction we will prove that for every node $v$, then every node $u$ in $N_{T}(v)/H[v]$ was already in the queue before $v$ if some node in between was already in the queue. In the first iteration we have that none of the nodes in $N_{T}(v)$ has passed through $Q$. Now, assume that at some iteration this hypothesis holds. If we take $v$ from $Q$, then line $6$ implies that we remove $v$ from $H[u]$ which means that every node in $N_{T}(u)/(H[u]-\{v\})$ already entered and left the queue maintaining this invariant. Coming back to our previous statement, we have that if $v$ left the queue there is a unique node $u = H[v]$ such that $u$ is in the path between $v$ and $w$. Otherwise $w$ would have left the queue before $v$. By Proposition~\ref{prop:prev-algo} we have then that $u$ is more important than $v$ in $T$. Finally, when $|Q| = 1$ at the beginning of one iteration, we have $v\in Q$ such that for every $u\in N_{T}(v)$ then $f(v, T_{v,u}) \geq f(u, T_{u,v})$ concluding that the last node in $Q$ is a root of $T$.
\end{proof}

\section{Proofs of Section~\ref{Sec:regularity}}
\subsection*{Proof of Proposition~\ref{Pr:ConsistencyChar}}

\begin{proof}
	$(\Rightarrow)$ If $C$ consistently roots trees, fix any tree $T$ and nodes $v,u$ in $T$ such that ${u,v}\in E(T)$ and $C(v,T) < C(u,T)$. Now, for a node $w$ in $T_{u,v}$ and node $w'$ not in $V(T)$, we will prove that $C(v, T+\{w,w'\})< C(u, T+\{w,w'\}).$ Since $C$ is a tree rooting measure, the fact that $C(v,T) < C(u,T)$ implies that $\maxcen(T)\subseteq T_{u,v}$. Set the node $r\in\maxcen(T)$ which is closer to $w$ in $T$. Now, by consistency we know that $\maxcen(T+\{w,w'\}) \subseteq \pi_{r, w'}\cup \maxcen(T) \subseteq V(T_{u,v})\cup \{w'\}$. Concluding by the rooting property of $C$ that $C(v, T+\{w,w'\})< C(u, T+\{w,w'\})$.
	
	\smallskip
	\noindent $(\Leftarrow)$ Suppose by contradiction there exists a tree $T$, nodes $r\in \maxcen(T)$ and $u\in V(T)$ such that for a node $w\not\in V(T)$ then there exists $r'$ in $\maxcen(T+\{u,w\})$ such that $r'$ is not in $\pi_{r,w}\cup \maxcen(T)$. Define the path $\pi_{r,w} :=z_1z_2...z_n$. Set $z^{\ast}\in \pi_{r,w}$ as the closest node to $r'$. Finally, set the path $\pi_{r'z^{\ast}} = s_1s_2...s_k$ connecting $z^{\ast}$ to $r'$. Now, we have that $s_1s_2...s_kz^{\ast}z_{i-1}...r = \pi_{r'r}$ is the path connecting $r'$ to $r$ in $T$. Then, since $r$ is a root in $T$ while $r'$ is not, we have that $C(z^{\ast}, T) > C(s_k, T)$. However, we have that $C(z^{\ast}, T+\{u,w\}) \leq C(s_k, T+\{u,w\})$ which contradicts monotonicity.
\end{proof}
\subsection*{Proof of Proposition~\ref{prop:monotonicity_potential}}
\begin{proof}
	$(\Rightarrow)$ Suppose by contradiction there exists a tree $T$, a subtree $T'\subseteq T$ and a node $v$ in $V(T)\cap V(T')$ such that $f(v, T') > f(v, T)$. Set a tree $H'$ and $v'\in V(H')$ such that $(v, T') \equiv (v', H')$ and $V(T')\cap V(H') = \emptyset$. In the the same way, set $H$ as a tree where $(v', H)\equiv (v, T)$ and $V(H)\cap V(T) = \emptyset$. In other words, $T$ is isomorphic to $H$ and $T'$ is isomorphic to $H'$ but they do not share nodes. Now, by close under isomorphism we have that $f(v, T') = f(v',H')$. On the other hand, there exists a sequence of edges as new leaves for $H'$, $e_1,e_2,...,e_n$ such that $H'+e_1+e_2+...+e_n = H$ which means that for some $i\in\{1,...,n\}$ it must be the case that $f(v', H'+ e_1+...+e_i) > f(v, T)$ but $f(v', H'+ e_1,...+e_{i+1}) \leq f(v, T)$. All of this implies that $C(v', H'+e_{1}+...+e_{i} + T + \{v,v'\}) > C(v,H'+e_{1}+...+e_{i} + T + \{v,v'\} )$ but $C(v', H'+e_{1}+...+e_{i+1} + T + \{v,v'\}) \leq C(v,H'+e_{1}+...+e_{i+1} + T + \{v,v'\})$ which means $C$ is not monotonic.
	
	\smallskip
	\noindent$(\Leftarrow)$ Take a tree $T$ and nodes $u,v$ in $T$ such that $\{u,v\}\in E(T)$ and $C(v, T) < C(u,T)$. Take nodes $w$ in $T_{u,v}$ and $w$ not in $T$. We have by potential functions definition that $f(v, T_{v,u}) < f(u, T_{u,v})$. Now, by hypothesis we have that $f(u, T_{u,v}) \leq f(u, T_{u,v} + \{w,w'\})$ since $T_{u,v}$ is a subtree of $ T_{u,v} + \{w,w'\}$. Adding this two facts together we have that $C(v, T+\{w,w'\}) < C(u, T+\{w,w'\})$ concluding that $C$ is monotonic.
\end{proof}

\subsection*{Proof of Theorem~\ref{prop:MonotonicSymmetric}}

\begin{proof}
	$(\Rightarrow)$ First, assume $u\not=v$ and set the path connecting $u$ and $v$ in $T$ as $\pi_{u,v} = w_1w_2...w_n$. Since $T'$ is a proper subtree of $T$, there exists a set of edges as leaves $e_1,e_2,...,e_k$ such that $T'+e_1+...+e_k=T_{u,w_{2}}$. Thus, by monotonicity and proposition~\ref{prop:monotonicity_potential} we have that $f(u,T')\leq f(u, T_{u,w_2})$. 
	By symmetry we have $f(u, T') \leq f(u, T_{u,w_2}) = f(u, T-T_{w_2,w_1})< f(w_{2}, T - T_{w_3,w_2})$. We can repeat this process until we reach $v$. We get then that $f(w_{n-1}, T-T_{w_n,w_{n-1}}) < f(w_n, T) = f(v, T)$ concluding the inequality when $u\not=v$. Now, if $u=v$ we have by monotonicity and proposition~\ref{prop:monotonicity_potential} that $f(u, T') \leq f(u, T)$. 
	
	\smallskip
	\noindent $(\Leftarrow)$ This direction is a direct application of lemma~\ref{Lemma:SymmetricPot} and proposition~\ref{prop:monotonicity_potential}.
\end{proof}

\section{Proofs of Section~\ref{Sec:alphabetagamma}}
\subsection*{Proof of Lemma~\ref{lemma:const}}

\begin{proof}
 To prove symmetry we will use lemma~\ref{Lemma:SymmetricPot}. Take any tree $T$ and two adjacent nodes $u,v$. We have that $\cpf(v,T) = l(\cpf(u,T_{u,v}))\mop \cpf(v, T_{v,u})$. We have that $l(\cpf(u,T_{u,v})) > \cpf(u,T_{u,v})$ by hypothesis. We also know that $(\mdom, \mop, \mone)$ is positively ordered and therefore, adding the fact that $\cpf(v, T_{v,u}) \geq \mone$ we conclude that $\cpf(v,T) > \cpf(u,T_{u,v}) \mop \cpf(v, T_{v,u}) \geq \cpf(u,T_{u,v})\mop \mone = \cpf(u,T_{u,v})$.
 
 We will prove that $\cpf$ is monotonic by exploding proposition~\ref{prop:monotonicity_potential}. In order to do this, we will prove that adding a leaf to the tree will not decrease the potential of $v$ after adding the leaf. Fix a tree $T$ and nodes $v,w$ in $V(T)$. For a node $h\not\in V(T)$ we will prove that $\cpf(v,T) \leq \cpf(v, T+\{w,h\})$. Set the path connecting $v$ to $w$ in $T$ as $\pi_{w,u} = z_1z_2...z_n$. Now, by definition of $\cpf$ we have that $\cpf(v,T) = \cpf(v,T_{v,w_2})\mop l(\cpf(z_2, T_{z_2,v}))$. Expanding this expression we have $\cpf(z_2, T_{z_2,v}) = f(z_2, T-T-T_{v,z_2} - T_{z_3,z_2})\mop l(\cpf(z_3, T_{z_3,z_2}))$. Adding up we got $\cpf(v,T) = \cpf(v,T_{v,w_2})\mop l(f(z_2, T-T-T_{v,z_2} - T_{z_3,z_2})\mop l(\cpf(z_3, T_{z_3,z_2})))$. Expanding until we reach $w$ we got $$\cpf(v, T)=\cpf(v,T_{v,w_2})\mop l(...\mop l(\cpf(w,T_{w,z_{n-1}}))...).$$
 
 Now, when we add the leaf $h$ we have $$\cpf(v, T+\{w,h\}) = \cpf(v,T_{v,w_2})\mop l(...\mop l(\cpf(w,T_{w,z_{n-1}})\mop l(\mone))...).$$
 
 Finally, the fact that $x<l(x)$ plus $(\mdom, \mop, \mone)$ is positively ordered, we got that $\cpf(w,T_{w,z_{n-1}})\mop l(\mone) > \cpf(w,T_{w,z_{n-1}})\mop \mone = \cpf(w,T_{w,z_{n-1}})$. Therefore,, $\cpf(v,T) > \cpf(v,T+h)$ by properties (1), (2) and (3).
 
 In order to finish the proof we must state the fact that, since $T'$ is a subgraph of $T$, there must exists a sequence of edges as leaves $e_1,...,e_k$ such that $T'+e_1+...+e_k = T$. Therefore, applying previous result for every $e_i$ we conclude that $\cpf(v, T')<\cpf(v, T)$ and $\cpf(v, T')=\cpf(v, T)$ just in the case that $T' = T$.
\end{proof}

\subsection*{Proof of Theorem~\ref{theo:const}}
\begin{proof}
	$(\Rightarrow)$ First we will prove that $x < \bar{\ell}(x)$. Take any tree $T$, node $v$ in $T$ and a leaf $w$ not in $V(T)$. By symmetry over potential functions we know that $\cpf(v,T) < \cpf(w, T+\{v,w\}) = \bar{\ell}(\cpf(v,T))$. Therefore, $\cpf$ satisfies (1).

	In second place, we will prove (2). Take any tree $T$ ,node $v$ in $T$ and a subtree $T'$ such that $v\in(T')$. Since $T'\subset T$ we have by monotonicity that $\cpf(v, T')\leq \cpf(v, T)$. Now, for a node $w$ not in $V(T')$, we have that $T'+\{w,v\}\subseteq T+\{w,v\}$. Thus, again by monotonicity we have that $\ell(\cpf(v, T'))=\cpf(w,T'+\{w,v\}) \leq \cpf(w,T+\{w,v\}) = \ell(\cpf(v, T))$. In other words, if $x\leq x'$ and $x$ is a subtree of $x'$ then $\ell(x)\leq\ell(x')$.
	
	Finally, we have to prove (3). Take a tree $T$ and a subtree $T'$. Now, fix a third tree $H$ and node $v$ such that $\{v\} = V(T')\cap V(H)$. We have by monotonicity and proposition~\ref{prop:monotonicity_potential} that $\cpf(v, T')\leq \cpf(v, T)$ and $\cpf(v, T)\mop \cpf(v,H)=\cpf(v, T+H) \geq \cpf(v, T'+H) = \cpf(v, T')\mop\cpf(v, H)$ since $T'+H$ is a subtree of $T+H$. In other words if $x$ is a subtree of $x'$ then $x\leq x'$ and $x\mop z \leq x'\mop z$ for $z\in\cpfran$ concluding the proof.
	
	\smallskip
	\noindent $(\Leftarrow)$ This proof is equivalent to the one of lemma~\ref{lemma:const} because we use subtrees in every case. 
\end{proof}

\subsection*{Proof of Proposition~\ref{Lemma:alphabetagamma}}

\begin{proof}
	We will use theorem~\ref{theo:const} from right to left. For (1), take $x\in\cpfran$. Then, $\ell(x) = ax+b \geq x+b > x$ for $a\geq 1$ and $b>0$. To prove (2), take $x\leq x'$ such that $x$ is a subtree of $x'$. Then, $\ell(x) = ax+b \leq ax'+b$. Finally, in order to prove (3) we first need to prove that for every $x\in \cpfran$ then $x\geq c$. First, we have that $\mone = c$. Now, by applying $\ell$, we got $\ell(\mone) = ac+b > c$ for $a\geq 1$ and $b>0$. Now, by structural induction, take any tree $T$ and node $v$ in $T$ such that $N_T(v) = \{w_1,...,w_n\}$ and assume $\cpf(w_i, T_{w_i,v}) \geq c$. We have that $\cpf(v, T) = \ell(\cpf(w_1, T_{w_1,v}))\mop\ell(\cpf(w_2, T_{w_2,v}))\mop...\mop\ell(\cpf(w_n, T_{w_n,v})) = \prod_{i=1}^{n}a(\cpf(w_i, T_{w_i,v}))+b \geq \displaystyle\frac{\prod_{i=1}^{n}ac+b}{c^{n-1}} > \displaystyle\frac{c^{n}}{c^{n-1}} = c.$ To conclude, take $x$ subtree of $x'$ such that $x\leq x'$. Take any $z\in \cpfran$. We have that $x\mop z = \displaystyle\frac{xz}{c} \leq \displaystyle\frac{x'z}{c} = x'\mop z$ since $\frac{z}{c} \geq 1$ because $z\geq c$ for every $z\in \cpfran$.
\end{proof}

\subsection*{Proof of Proposition~\ref{prop:InfiniteRooting}}

\begin{proof}
	For this purpose we will just study how $f_{\mop_{c},\ell_{a,b}}$ assign values to the star graph with $n$ nodes, $S_n$, and the line graph with $m$ nodes, $L_{m}$. In one side, we have that $f_{\mop_{c},\ell_{a,b}}(0, S_n) = \displaystyle\frac{(ac+b)^{n}}{c^{n-1}}$. On the other hand, for the line we have that $f_{\mop_{c},\ell_{a,b}}(0, L_m) = \ell^{m}(c) = a\ell^{m-1}(c) + b = ... = a^{m}c +\sum_{i=0}^{m-1}a^ib$. Now, fix $c=1$ and we have that $f_{\mop_{c},\ell_{a,b}}(0, S^{n})< f_{\mop_{c},\ell_{a,b}}(0, L_{m})$ if and only if $ \displaystyle(a+b)^{n} < a^{m} +\sum_{i=0}^{m-1}a^ib$. Now, for a fixed pair $a,b$, and $n\geq 1$ take $m(a,b,n)=\arg\min_{m\geq 1} \{(a+b)^{n} < a^{m} + \sum_{i=0}^{m-1} a^{i}b\}$. Since $a^{m} + \sum_{i=0}^{m-1} a^{i}b$ is unbounded in $m$ for $a \geq 1$ and $b>0$ we have that $m(a,b,n)\in\mathbb{N}$. Finally, we have that $(a+b)^n \geq a^{m(a,b,n)-1} + \sum_{i = 0}^{m(a,b,n)-2}a^{i}b$. Therefore, we have that in the tree $S_{n} + L_{m(a,b,n)-1}$ the root is in $S_{n}$ while in $S_{n}+ L_{m(a,b,n)}$ the root for $f_{\mop_{c},\ell_{a,b}}$ is in $L_{m(a,b,n)}$. On the other hand, the values of $f_{\mop_{c},\ell_{a,b}}(0, S_n)$ and $f_{\mop_{c},\ell_{a,b}}(0, L_m)$ are also unbounded for $a\geq 1$ and $b>0$. In other words, for every value $n\geq 4$ there exists $a,b$ and value $m\geq n$ such that $m(a,b,n) = m$ concluding that there exists an infinite amount of potential functions which differ in the ranking of $S_n + L_{m(a,b)-1}$ and $S_n + L_{m(a,b,n)}$ for every choice of $n\geq 4$.
\end{proof}

%

\end{document}